\documentclass[sn-mathphys-num]{sn-jnl}

\usepackage{graphicx}%
\usepackage{multirow}%
\usepackage{amsmath,amssymb,amsfonts}%
\usepackage{amsthm}%
\usepackage{mathrsfs}%
\usepackage[title]{appendix}%
\usepackage{xcolor}%
\usepackage{textcomp}%
\usepackage{manyfoot}%
\usepackage{booktabs}%
\usepackage{algorithm}%
\usepackage{algorithmicx}%
\usepackage{algpseudocode}%
\usepackage{listings}%
\usepackage{braket}%

\begin{document}

\title[Quantum simulations of complex systems]{Quantum simulations of complex systems}

\author[1,2]{\fnm{Oliver} \sur{Morsch}}\email{morsch@df.unipi.it}

\author[3]{\fnm{G. Massimo} \sur{Palma}}\email{massimo.palma@unipa.it}

\author[2,4]{\fnm{Davide} \sur{Rossini}}\email{davide.rossini@unipi.it}

\affil[1]{\orgdiv{CNR-INO}, \orgaddress{\street{via Moruzzi 1}, \city{Pisa}, \postcode{56124}, \country{Italy}}}

\affil[2]{\orgdiv{Dipartimento di Fisica}, \orgname{Università di Pisa}, \orgaddress{\street{Largo B. Pontecorvo 3}, \city{Pisa}, \postcode{56127} \country{Italy}}}

\affil[3]{\orgdiv{Dipartimento di Fisica e Chimica - Emilio Segre'}, \orgname{Università di Palermo}, \orgaddress{\street{via Archirafi 34}, \city{Palermo}, \postcode{90123},  \country{Italy}}}

\affil[4]{\orgdiv{INFN, Sezione di Pisa}, \orgaddress{\street{Largo B. Pontecorvo 3}, \city{Pisa}, \postcode{56127} \country{Italy}}}

\abstract{In this review we give a brief overview of quantum simulation as applied to the study of complex systems. In particular, we cover the basic ideas of quantum simulation, neuromorphic computation, the Sachdev-Ye-Kitaev model, as well as applications to quantum batteries.}

\keywords{Quantum simulations, Complex systems, Neural networks}

\maketitle

\section{Introduction} 

Quantum simulators are one of the leading characters of the modern quest for quantum information processing paradigms. Strongly related but at the same time different, both conceptually and experimentally, from quantum computers, they are receiving a stronger and stronger interest after the advent of the noisy intermediate scale quantum regime of quantum devices. The key reason for such interest resides in the fact that quantum simulators, often referred to as analogue quantum computers, are specialized quantum devices that mimic the behavior of complex quantum systems, while being easier to manipulate. Unlike universal quantum computers, these simulators are designed to address specific problems and follow the mathematical formalism of the quantum systems they are intended to emulate. Although they cannot perform general-purpose computation, quantum simulators are expected to solve suitable problems earlier than digital quantum computers, because they have less stringent requirements on the qubits quality.

This article is far from being a full review of the current state of the art in quantum simulation, for which an incredibly increasing amount of literature has been produced in the last decades~\cite{NielsenChuang, Buluta2009, Ladd2010, Kendon2010, saff10, Cirac2012, Hauke2012, Schaetz2013, Georgescu2014, Gross2017, Wendin2017, flam19, Bruze19, kajer20, monroe21, Altman2021, Bauer2023_B, bauer23, rama23, fause24}.
In contrast, it has to be rather intended as an idiosyncratic introduction to the key features of quantum simulators and to their capability to address complex systems, reflecting our personal interests (and cultural bias). We have however tried to provide a logical sequence of topics to guide the reader through the main motivations behind quantum simulators, to its theoretical aspects and to its physical implementations.

To this end, the work is conceptually divided into three separate parts. We start, in Sec.~\ref{sec:background}, with some historical background and put quantum simulators in perspective, highlighting its similarities and its differences, and most importantly its advantages, with respect to ``standard" quantum computing. In Sec.~\ref{sec:neuromorphic}, we illustrate the links between quantum simulators and quantum neural networks, and in particular with quantum reservoir computing, a basic yet very effective quantum machine learning protocol. Then, in Sec.~\ref{Kitaev}~and~\ref{usefulSYK} we focus on a paradigmatic complex system, the Sachdev-Ye-Kitaev (SYK) model. As this is not meant as comprehensive review of all the possible complex systems which might implement a quantum simulators, we have singled out the SYK model for its reach features and cross-sectoral interest, from condensed-matter to high-energy physics communities.
These make it particularly suited in the description of the key features of good quantum simulators and, more in general, of their relevance for quantum technologies.
In the last part, in Sec.~\ref{experimental}, we review the physical platforms which are currently more promising for the implementation of quantum simulators. 
Finally, in Sec.~\ref{conclusions}, we summarize and draw our final remarks.

\section{Quantum Simulators vs Quantum Computers}
\label{sec:background}

The history of the concepts of quantum computation and of quantum simulation are strongly intertwined. The scientific lore attributes the birth of  quantum simulators to the seminal Richard Feynman 1982 paper~\cite{Feynman1982467}, however the true beginning for a more complete and correct narrative would be the 1981 Physics of Computation Conference,
jointly organized by MIT and IBM at MIT’s Endicott House. The conference picture is, for today community of quantum computation scientists, nearly as iconic as the Solvay 1927 conference picture is for quantum physicists. 
The list of participants included, among the others, Freeman Dyson, John Wheeler, Gregory Chaitin, Konrad Zuse, Rolf Landauer, Paul Benioff, Edward Friedkin, Lev Levitin, Daniel Greenberger, Charles Bennett (the picture is shot by him), and Richard Feynman. The hot topic then was logical vs thermodynamical reversibility, i.e., the intrinsic, unavoidable energetic cost of computation. Tom Toffoli and Edward Friedkin had developed the reversible gate model of computation, while Paul Benioff had developed a reversible quantum Turing machine. Following Rolf Landauer's motto ``information is physical", the interest moved then towards atomic scale implementations of reversible computation. It should be stressed that these were atomic scale {\em classical} computers, e.g., atomic scale computing devices performing classical computation. Any quantum effect was then considered as a source of noise~\cite{Toffoli1980632, Benioff19821581, Benioff1982515}. 
Following this conference, Feynman, in his above-mentioned paper~\cite{Feynman1982467}, pointed out that simulating quantum systems ---especially large multipartite quantum systems--- is hard  on a classical computer, due to the exponential growth of their Hilbert space. He suggested that a better strategy to simulate a complex quantum systems would be to map its dynamics into the controlled dynamics of a different quantum system, thus introducing the concept of a {\em quantum simulator}. 

Shortly after, David Deutsch gave birth to the quantum computation paradigm as we know it, with his seminal papers on universal quantum Turing machines and on universal quantum gates, introducing therefore the concept of {\em quantum algorithms}~\cite{Deutsch198597, Deutsch199847}. Here we are not going to discuss in detail the power of quantum algorithms, but let us stress that both quantum Turing machines and quantum gate based computations imply a sequence of controlled, clocked, interactions between quantum subsystems, all of the same dimensions (the qubits). Quantum computers are universal, i.e., they can perform any computational task ---any algorithm. Of course a quantum computer can simulate any quantum system. However, on the one hand, this is different from Feynman's concept of quantum simulators and, on the other hand, it can be demanding in terms of quantum gates. The reason being that often the individual subsystems of the multipartite system one wants to simulate are not necessarily bidimensional, i.e., they are not necessarily qubits, therefore one would need several qubits to encode each single subsystem. Furthermore the global evolution of the system must be decomposed in terms of a sequence of single and pairwise gates.

The problem of the resources (the number of quantum gates) needed to simulate quantum systems is nowadays particularly relevant, given the development, in an astonishingly short time, of present-day quantum computers, which have reached the so called Noisy Intermediate Scale Quantum (NISQ) regime. Such regime is still far from being quantum fault tolerant and it is indeed noisy. Although great progress has been made in building quantum computers with a larger and larger number of qubits, the precise control of their quantum evolution required to  implement complex quantum algorithms is still not satisfactory. This has revamped the interest for {\em quantum simulators}, which, in some of their versions, can be thought as quantum analogue computers. They are particularly convenient whenever the digital quantum simulation of a quantum system on a digital quantum computer ---i.e., the simulation of a quantum system in which its state is encoded into a string of qubits and its dynamics is mapped into a sequence of gates--- is demanding in terms of number of gates and of qubits. In fact, one of the advantages of quantum simulators is that they do not necessarily require a register of interacting qubits, but they can consist of interacting subsystems living in Hilbert spaces of arbitrary dimensions (qutrits, continuous variables, etc). Furthermore, although they are not universal their implementation is, in general, less demanding than building a full, universal, quantum computer. It should be also mentioned that they are typically less prone to the disruptive effect of quantum noise. Summing up, quantum simulators leverage on the intrinsic quantum complexity of multipartite interacting quantum system to achieve specific computational advantage with respect to both  universal classical and quantum digital computers.
The typical scenario in which quantum simulators can prove to be a powerful tool is when the original system is difficult to manipulate, when one is interested in analyzing toy models with complex Hamiltonian and, in a more applied scenario, for a quantitative analysis of strongly interacting fermionic systems, like the ones encountered in quantum chemistry of in material science.

More recently, complex quantum systems have been used not only to simulate strongly interacting systems, but also to simulate and characterize open system dynamics. In particular, complex systems like bosonic quantum networks can be engineered to mimic and probe structured environments~\cite{Nokkala_2016, GarciaPerez_2020,  Renault_2023, Patsch_2020}, often characterized by a non-Markovian open dynamics. The characterization of complex networks can be performed efficiently by random quantum walks~\cite{Magano_2023}. An extensive discussion of quantum random walks~\cite{Kempe01072003} is outside the scope of this review.

\subsection{Quantum annealing}

So far we have discussed the conceptual differences between quantum computing and quantum simulation. Let us now show how quantum simulators can be used to implement non algorithmic (non deterministic) computation i.e. how they can be used to implement heuristic algorithms. During the same years in which the concepts of quantum simulation and quantum computation evolved, a strong interest for complex system in general and for spin glasses in particular has emerged in the field of statistical mechanics (the Nobel prize awarded to Giorgio Parisi in 2021 and to John Hopfield in 2024 are an acknowledgment of the relevance of the field).  Spin glasses are disordered frustrated systems characterized by a complex fractal energy landscape. It became soon clear that the same techniques used to find the ground state of the system could be uses to solve heuristically combinatorial or optimization problems. Such technique, known as {\em Simulated Annealing}~\cite{Mezard20091}, is a probabilistic algorithm to search for the global minimum in a complex energy landscape. The basic idea is map the function whose minimum one is looking for into the free energy of a suitable complex system. The system explores its energy landscape and, by slowly lowering the temperature (hence the name annealing), it will end, with high probability, into its lowest energy state. This technique finds useful application when one is searching for the lowest of many discrete minima in a large search space, a scenario encountered, e.g  in the traveling salesman problem, the boolean satisfiability problem, protein structure prediction, and scheduling. 

Shortly after, the quantum version of this process, namely {\em Quantum Annealing} (QA)~\cite{Farhi2001, Zecchina, Mezard20091, Rajak2023}, was put forward with the role of thermal fluctuations being played by quantum fluctuations. The basic idea is that, while in Simulated Annealing finite temperature fluctuation prevent the system from being trapped into a local minimum, in QA the same mechanism is played by tunneling. To illustrate the idea suppose that the solution of a combinatorial problem can be mapped into the ground state of a many body Hamiltonian $H_c$, typically a Ising-like Hamiltonian which can be expressed in terms of the Pauli $\{ \sigma^z \}$ operators and which can be viewed as a potential energy term. To let the system evolve towards the ground state of $H_c$, one introduces a driving Hamiltonian $H_d$, i.e., a ``kinetic" terms which, for an Ising model, corresponds to a local transverse field term $-h \sum_i \sigma^x_i$. Such kinetic term does not commute with $H_c$ and its eigenstates are like delocalized ``plane waves" in the $\sigma^z$ basis. The QA protocol relies on the adiabatic theorem ---hence the name Adiabatic Quantum Computation (AQC) often used instead of QA --- to drive the system from the ground state of $H_d$ to the ground state of $H_c$. To do so, one engineers a time dependent quantum adiabatic Hamiltonian 
\begin{equation}
H_{qa}(t) = \alpha (t) H_c + \beta (t) H_d
\end{equation}
where $\alpha (t)$ and $\beta (t)$ are slowly varying function such that $\alpha (0) \ll \beta (0)$ and $\alpha (T) \gg \beta (T)$, where $t=0$ and $t = T$ are the initial and the final state of the protocol. If the process is done adiabatically and if the system is in the ground state of $H_d$ at time $t=0$, it will remain in the instantaneous ground state of $H_{qa}$ and, at $t=T$, it will be in the ground state of $H_c$, which encodes the desired solution of the optimization problem. Let us recall that the adiabaticity condition holds as long as
\begin{equation}
\frac{\max {\mathcal R(t)}}{\min [\Delta (t)]^2} \ll 1
\end{equation}
where ${\mathcal R(t)} =  \langle \varphi_1 | (d H_{qa}/dt) | g(t)\rangle $ is the rate of change of the $H_{qa}$ between the instantaneous ground state $|g(t)\rangle$ and the first excited state $|\varphi_1\rangle$, and $\Delta (t)$ is the corresponding energy gap. The $\max$ and $\min$ being evaluated as a function of time.

There is evidence that classical simulation of QA is more efficient than simulated annealing~\cite{Farhi2001, Zecchina}, with quantum tunneling being more efficient than thermal fluctuations in jumping across thin and high potential barriers.  Its true power emerges however when, in adherence with the concept of quantum simulation, it is implemented on a true quantum hardware, where tunneling is not classically simulated but occurs intrinsically in the hardware dynamics. The advantage with respect to digital quantum computers is also evident, as no exact control of a sequence of gates and no error correction protocol needs to be implemented.

\section{From quantum simulators to quantum neuromorphic computation}
\label{sec:neuromorphic}

The development of classical neuromorphic computation is prompting a strong interest for its quantum counterpart. This is a substantial change in the approach with which the dynamics of a complex system can be used to perform a computational task. Machine learning is a non-algorithmic form of computation. It can of course be run on a digital computer, but there is a growing interest towards its implementation on non conventional dedicated hardware, ranging from mechanical systems to chemical reactions. Out of the several quantum machine learning schemes, in this review we focus our attention to {\em quantum extreme learning machines}(QELM) and {\em quantum reservoir computing} (QRC)~\cite{martinuzzi2020blog, nakajima2019boosting,mujal2021opportunities,martinezpena2020information, Innocenti2023, Suprano2024}. The reason for this choice is twofold. First of all, this is perhaps the simplest ---both classical and quantum--- non algorithmic machine learning scheme. Second, it can be implemented into a broad range of different platforms and hardly requires any control. Reservoir computers and their memoryless counterparts, as extreme learning machines (ELM), are a simplified version of recurrent neural networks with a training process only at the output stage.

\begin{figure}
  \begin{centering}
    \includegraphics[width=12 cm]{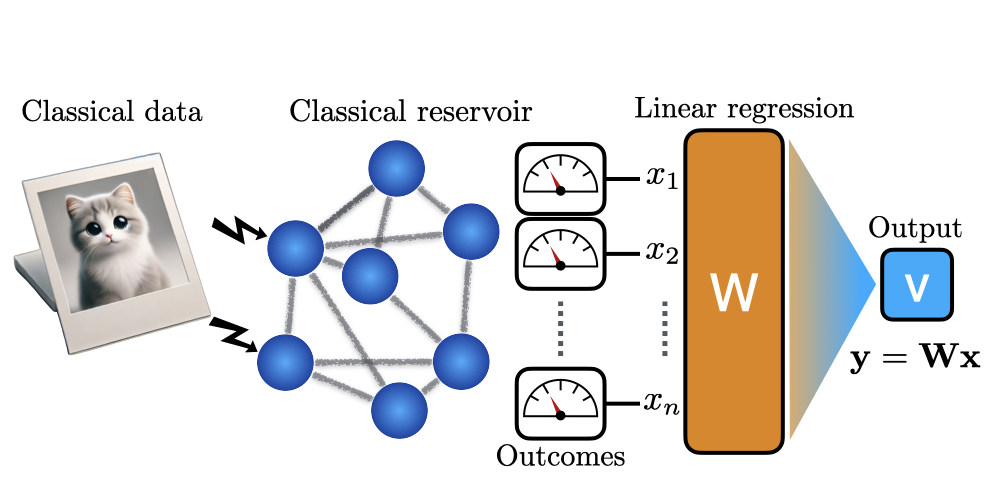}   
    \caption{Classical reservoir computing: classical data are fed into a reservoir consisting of a network of classical interacting subsystems. After the information about the input propagates into the reservoir, a (sub)set of nodes is measured. A linear regression is then made on the measurement outcomes to fit the desired function of the input. In the above example, the system is trained to tell if the input picture is a cat or not.}
    \label{CRC}
  \end{centering}
\end{figure}

Let us first describe the working of a classical reservoir computer. A set of input signals ${\bf x}_k$, where $k = 1, \ldots, M_t$ labels the different training inputs, is fed into a reservoir, i.e., a system consisting of a large number of subsystems randomly coupled with each other (see Fig.~\ref{CRC}). The strength of the couplings is fixed and it is not trained. The randomness of the coupling within the reservoir guarantees the absence of symmetries which would restrict the phase space of its internal states accessible during its evolution. Different input signals will trigger different reservoir dynamics. The role of the reservoir is twofold: it expands the phase space of the dynamics induced by the signal and it provides the nonlinearity required to perform arbitrary function evaluations. The reservoir subsystems can be anything, from mechanical systems to photonic platforms, to chemical elements. At a fixed time, the state of a subset of the reservoir subsystems is measured, the measurement outcome being a function $f({\bf x}_k)$ of the input signal. The measurement outcomes are then linearly combined through suitable trained weights $W$, so that the $Wf({\bf x}_k) ={\bf y}_k$ for all $k$ where ${\bf y}_k$ is the desired output. Such scheme is much simpler than standard deep learning of recurrent networks, where the training takes place also at the level of the interlayer couplings. In general, reservoir computers will have a memory, which is useful when one is interested in tasks like time sequence recognition. If the state of the reservoir is reset after each input, we have an ELM, particularly useful in tasks like classification. 

\begin{figure}
  \begin{centering}
    \includegraphics[width=12 cm]{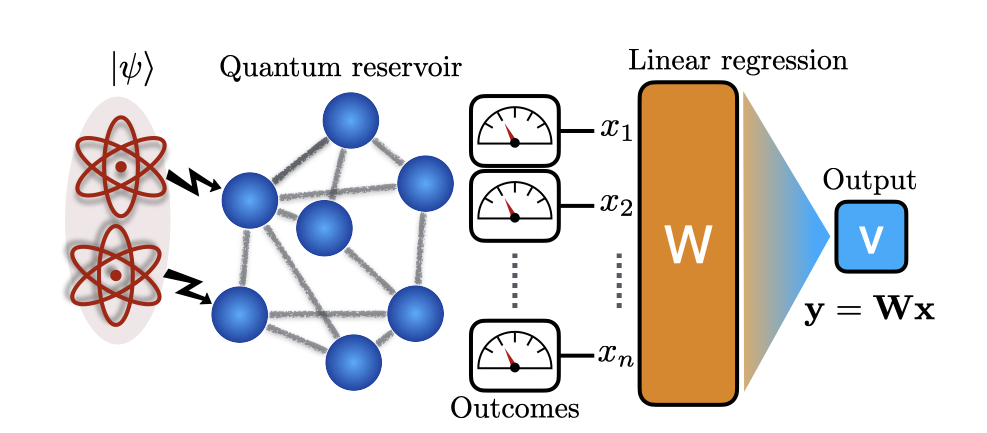}   
    \caption{In the quantum scenario an input is fed into a reservoir consisting of a multipartite quantum system. A measurement is performed on some, or all, the reservoir subsystems. The training is performed on the measurement output probabilities. The task can be either purely quantum (e.g., to reconstruct the input state) or classical (e.g., to process some classical data encoded in the input quantum state).}
    \label{QRC}
  \end{centering}
\end{figure}

In QRC and in QELM, the information is ``processed" by a  quantum reservoir, again consisting of randomly coupled quantum subsystems (see Fig.~\ref{QRC}). Also in the quantum case, the randomness in the internal couplings guarantees that symmetries do not confine the reservoir dynamics prompted by external systems into a small subspace. 
The input  signal of QRC and QELM can be either classical or quantum, i.e., one can encode the input information into a classical signal like a strong laser pulse or an impulse in a nanomechanical array, or couple a quantum state (i.e., a spin) in a given state to a multipartite quantum system. Classical input states can encode classical information, while quantum input states can be used to encode either classical or quantum information. In other words, QRC and QELM can either perform efficiently classical tasks like quantitative predictions of some quantity, or even quantum tasks, like quantum state tomography or quantum entanglement witnessing. 

In classical ELM, the training dataset is $\{ ({\bf x }_k, {\bf y}_k)\}_k$, the model to train is ${\bf x}\mapsto Wf({\bf x}$), and $W$ is trained to minimize the cost function $\| {\bf y}_k -  Wf({\bf x}_k) \|^2$ for all $k$. In QELM, the input are the training  states $\varrho_k$. Once the input state is coupled to the reservoir, they will jointly evolve according to a unitary evolution followed by a measurement. This amounts to evolving the input state with a completely positive trace preserving quantum map $\Lambda $ and then performing  a positive operator valued measurement (POVM) $\mu_b$ (here $b$ runs over the possible measurement outcomes)  on the evolved input state $\Lambda (\varrho_k)$, the probability measurement outcomes being ${\bf p}_{\Lambda , \mu} (\varrho )_b = Tr [ \mu_b \Lambda (\varrho ) ]$. The QELM training datasets are then $\{ ({\varrho }_k, {\bf y}_k)\}_k$, the model to train is ${\varrho }\mapsto W{\bf p}_{\Lambda , \mu} (\varrho )$ and again $W$ is trained to minimize the cost function $\| {\bf y}_k -  W{\bf p}_{\Lambda , \mu} (\varrho_k ) \|^2$ for all $k$. It is important to note that, while in classical ELM the function $f({\bf x}_k)$ (i.e., the dependence of the measurement outcomes on the input) is non linear, in QELM the measurement outcome probabilities ${\bf p}_{\Lambda , \mu} (\varrho )_b $ depend linearly on the input state $\varrho_k$. This implies that either the nonlinearity is in the encoding of some classical parameter in the input state or, if one is interested in learning some non linear functional of the input state, one resorts to consecutive multiple injections of $\varrho_k$. An interesting feature of QELM is that recovery of information about the input cane be recovered even after the onset of scrambling \cite{Vetrano2025} (more on scrambling in \ref{scrambling}).

From the above discussion, it is clear how QELM can turn complex quantum systems into effective quantum computing devices, somehow interpolating between quantum simulators and quantum computers, with a clear advantage in terms of quantum resources needed, of noise resilience, and of fault tolerance, requiring very little quantum control. Below we provide two different examples of implementation of QELM on two completely different physical platforms.

As first example, we illustrate how a QELM can predict molecular energy surfaces. The molecular energy surface depends on the molecular geometry, parametrized by generalized coordinates. For example, the geometry of a water molecule is fully determined by the two O-H bonds lengths $r_1$ and $r_2$ and by their relative angle $\varphi$. 
The QELM has been implemented on a true quantum computer \cite{LoMonaco2024}, the IBM BRISBANE quantum processor, a device with 127 superconducting transmon qubits). The first step is to encode the input data in a quantum state. As this is done by suitable rotations along the qubit $z$ axis, the generalized coordinates are mapped into angles by rescaling each bond length with respect to a reference one $\overline{r}$ (for water $\overline{r}_{{\rm H}_2{\rm O}}=2$\AA). The bond angles are instead rescaled by a factor $2$. The input data of water are thus, for example,  of the following form:
\begin{equation}
    \boldsymbol{x}=\left\{\frac{r_1}{\overline{r}}\pi\,,\,\frac{r_2}{\overline{r}}\pi\,,\,\frac{\varphi}{2}\right\}\,.
\end{equation} 
The information scrambling of the input information into the reservoir is implemented by a random operator $W$ acting on the $N$ reservoir qubits. The overall encoding of the $ x_1 \ldots x_{\mathcal{X}}$ input parameters plus the mixing is therefore described by a product of local rotations by an angle $x_i$ along the $z$ axis of the $N$  reservoir qubits interspersed by joint random unitaries on the $N$ reservoir qubits:
\begin{equation}
\label{eq:Fourier_Embedding}
    U(\boldsymbol x)=e^{-x_{\mathcal{X}}G}W^{(\mathcal{X})}\dots W^{(2)}e^{-x_1G}W^{(1)} ,
\end{equation}
where 
\begin{equation}
\label{eq:G_H2O}
G=\frac{1}{2}\sum_{i=1}^{N}\sigma_i^z .
\end{equation}
All the $N$ reservoir qubits are then measured in the $\sigma^z$ basis and the training of probabilities is made as described above. The predicted energies have an accuracy of the order of $10^{-2} - 10^{-3}$ Hartree.

As a second task, which has been experimentally implemented on a QELM, we describe the state estimation of a single photon polarization. Here the reservoir dynamics consists of a coined quantum walk (QW) in polarization and orbital angular momentum (OAM) of single photons~\cite{Suprano2024}.

The expectation values of observables on the input polarization states are obtained by using the reservoir dynamics to transfer this information into the larger OAM space that is then measured. The joint polarization / OAM dynamics is a sequence of unitary operators $C_k = {\mathcal I}_{o}\otimes{\mathcal U}_k$ acting only on the polarization degrees of freedom (${\mathcal U}_k$ is the so called coin operator, in the quantum random walk jargon and ${\mathcal I}_{o}$ is the identity on the OAM) and of conditional transitions ${\mathcal T}$ on the OAM degrees of freedom with 
\begin{equation}
{\mathcal T} = \sqrt{p} \, {\mathcal I}_o\otimes{\mathcal I}_p + \sqrt{1-p} \, \big( {\mathcal E}_+ \otimes |\! \downarrow \rangle \langle \uparrow \! |
+ {\mathcal E}_- \otimes |\!\uparrow \rangle \langle \downarrow\! | \, \big),
\end{equation}
where $|\!\uparrow \rangle$ and $|\!\downarrow \rangle$ are the polarization states and ${\mathcal E}_{\pm } |j\rangle = |j\pm 1 \rangle$, where the OAM states $|j\rangle$ play the role of the walker position state. 
The joint polarization /OAM state after $n$ steps will be $|\Psi_n \rangle = (\Pi_k^n {\mathcal T}{\mathcal C}_k )|0,\psi\rangle$ where $|\psi\rangle$ is the initial polarization state. After $n$ steps, a projective measurement on the OAM degrees of freedom is performed and the probability outcomes are trained to infer the expectation values of operators on the polarization state 
$|\psi\rangle$.

\section{Quantum simulation of a paradigmatic complex system}
\label{Kitaev}

As already mentioned in the introductory part of this review paper, while it is not our purpose to provide an exhaustive treatment of the field of quantum simulation, 
we feel it would be instructive to focus on a specific, yet paradigmatic, example of a complex system where the above concepts can be applied.
In fact, a large variety of quantum networks have been shown to exhibit complex features, such as the presence of entanglement, therefore they constitute an optimal testbed to study quantum communication and computation schemes, as well as genuine features of quantum dynamics~\cite{NielsenChuang, Acin_2007, Cuquet_2009, Perseguers_2010, Biamonte_2019}.

Below we provide the reader with a general idea of what can be done in this context, focusing on a prototypical example of a complex quantum system, which exhibits a rich many-body phenomenology and is receiving considerable interest in the last decade: namely, we consider the SYK model~\cite{Sachdev-Ye_1993, Kitaev_2015}. In this section we briefly discuss why it is considered to be crucial to find a way to efficiently simulate it, and then provide some details on its most relevant physical properties.
Then, in Sec.~\ref{usefulSYK}, we will address possible a technological application which could be able, in principle, to exploit the quantum advantage associated to the genuinely quantum features of its complex dynamics.

\subsection{The Sachdev-Ye-Kitaev model}
\label{sec:SYK_model}

The SYK model was introduced in condensed-matter physics as a prototype of strange metal, with a non-Fermi liquid behavior. 
Namely, it describes $Q$ strongly coupled spinless fermions, moving on a $N$-site lattice, and characterized by random and uncorrelated two-particle interactions of arbitrary strength~\cite{Sachdev-Ye_1993, Kitaev_2015}.
One of its basic features is the lack of quasiparticle excitations, resulting in an exponentially dense low-energy spectrum (Sec.~\ref{sec:SYK_quasiparticle}). 
Besides that, at large $N$ it is possible to obtain an exact formal treatment (Sec.~\ref{sec:SYK-LargeN}), which reveals that the SYK model is maximally chaotic: more precisely, it displays fast scrambling, with a nonzero entropy density at vanishing temperature, and exhibits a volume-law bipartite entanglement entropy for all the eigenstates at any energy scale (even for the ground state).
These facts, together with the emergence of a conformal symmetry at low temperatures, suggest that it has an interesting holographic dual which describes a theory of quantum gravity~\cite{Rosenhaus_rev, SYK_rev, Sachdev_rev}.

The Hamiltonian reads
\begin{equation}
  H_{\rm SYK} = \frac{1}{(2 N)^{3/2}} \sum_{i,j,k,l=1}^N \tilde J_{ijkl} \, c^\dagger_i c^\dagger_j c_k c_l
  - \mu \sum_{i=1}^N c^\dagger_i c_i .
  \label{eq:SYK_ham}
\end{equation}
where $c_i^{(\dagger)}$ are anticommuting operators in the second quantization formalism, which denote annihilation (creation) operators for spinless fermions moving between $N$ different sites of a lattice,
labeled by $i= 1,\ldots,N$.
The couplings $\tilde J_{ijkl}$ are independent Gaussian distributed complex variables
with zero average $\langle \! \langle \tilde J_{ijkl} \rangle \! \rangle = 0$ and variance
$\langle \! \langle |\tilde J_{ijkl}|^2 \rangle \! \rangle = J^2$ ($J \in \mathbb{R}$),
which satisfy $\tilde J_{ijkl} = - \tilde J_{jikl} = - \tilde J_{ijlk} = \tilde J^*_{klil}$,
while $\mu$ acts as a chemical potential.
The $1/\sqrt{N^3}$ scaling in front of the interaction strength guarantees that the system bandwidth 
is of the order of $N$, in the $N \to \infty$ thermodynamic limit, such that extensivity
of thermodynamic quantities as the energy is preserved.
Hereafter we redefine $J_{ijkl} = \tilde J_{ijkl} / (2N)^{3/2}$.

The SYK model conserves the total number of fermions, as
the Hamiltonian $H_{\rm SYK}$ commutes with the operator
\begin{equation}
  {\cal Q} = \sum_{i=1}^N c^\dagger_i c_i ;
\end{equation}
the second contribution in Eq.~\eqref{eq:SYK_ham} ensures that, in the grand canonical formalism,
the average density of fermions on each site is $Q/N \equiv \langle {\cal Q} \rangle/N \in [0,1]$.
As we shall see below, the situation for which $Q/N = 1/2$ is of particular importance and is
referred to as the {\em half-filling} condition.
In the following, we will be mostly interested in quantities or observables $\cal{O}$ averaged
over the distribution of the couplings $\{ J_{ijkl} \}$.
Thus we adopt the notation
\begin{equation}
  \langle \! \langle {\cal O} \rangle \! \rangle \equiv
  \int P(\{ J_{ijkl} \}) {\cal O} ( \{ J_{ijkl} \} ) \, {\rm d} \{ J_{ijkl} \},
\end{equation}
where $P(\{ J_{ijkl} \})$ denotes the corresponding probability distribution of the couplings.

Finding a way to efficiently simulate Eq.~\eqref{eq:SYK_ham} is of the uttermost importance, as currently the SYK model lacks of a full analytic treatment and several issues related to its emerging physics are still unsolved.
The main obstacle for classical simulations is the exponential growth of the Hilbert space with the number of sites, a fact that drastically limits the application of exact-diagonalization (ED) methods to systems with $N \sim \mathcal{O}(10)$.
As an alternative, several quantum hardware platforms have been proposed and thoroughly discussed, as clarified below (Sec.~\ref{subsec:SYK_simul}).

\subsection{Simulations of the SYK model}
\label{subsec:SYK_simul}

In general, any numerical simulation of a quantum many-body system on a lattice with arbitrary geometry has to face with the fact that the global Hilbert space $\mathcal{H}$ has an exponentially growing dimension, ${\rm dim} (\mathcal{H}) = 2^N$.
This is true also for the SYK model, which lacks of any particular symmetry (apart from the conservation of the number of fermions) and for which no efficient classical numerical method is known yet.
In fact, for the $H_{\rm SYK}$ Hamiltonian in Eq.~\eqref{eq:SYK_ham}, $\mathcal{H}$ can be decomposed as a direct sum of subspaces with a fixed number $Q = 1,2, \ldots, N$
of fermions:
\begin{equation}
  \mathcal{H} = \bigoplus_{Q=1}^{N} \,\mathcal{H}_{Q}, \qquad \mbox{with} \;\; 
  \operatorname{dim} (\mathcal{H}_Q) = \binom{N}{Q} .
\end{equation}
In the thermodynamic limit $N \to \infty$, the only relevant subspace is the one corresponding to the half-filling condition $Q = N /2$, whose cardinality is the largest and keeps scaling exponentially with $N$.
It is thus useful to concentrate on $\mathcal{H}_{N/2}$: its dimension keeps growing exponentially with $N$, although for
$N \sim 10$ its computational complexity is one order of magnitude smaller than that for $\mathcal{H}$.

In practice, it is convenient to transform the SYK model into a system of interacting spin-$1/2$
particles, through a Jordan-Wigner (JW) transformation~\cite{Solano_2017}.
This enables us to map creation/annihilation operators for fermions $c_j^{(\dagger)}$
into spin-$1/2$ Pauli matrices $\sigma^\alpha_j$ ($\alpha = x,y,z$) and the corresponding raising/lowering operators
$\sigma^\pm_j = \tfrac12 (\sigma^x_j \pm \sigma^y_j)$.
To correctly transform fermionic anticommutation relations into commutation relations for the Pauli matrices,
one needs to exploit the so-called JW string, in such a way that:
\begin{equation}
    c^\dagger_i = \sigma^+_i \bigg( \prod_{j<i} \sigma_j^z \bigg)
\end{equation}
The JW string delocalizes the single excitation on the whole spin network.
In the case (i) where all four indexes $i,j,k,l$ in Eq.~\eqref{eq:SYK_ham} are different
($i\neq j\neq k \neq l$), the two-body interaction term in $H_{\rm SYK}$ is mapped into
\begin{subequations}
\begin{equation}
  c_i^\dagger c_j^\dagger c_k^{\phantom{\dagger}} c_l^{\phantom{\dagger}} = 
  \operatorname{sgn}(i-j)\operatorname{sgn}(k-l)
  \prod_{\zeta=\zeta_1+1}^{\zeta_2-1} \sigma^z_\zeta
  \prod_{\zeta'=\zeta_3+1}^{\zeta_4-1} \sigma^z_{\zeta'}\,\,
  \sigma^+_i\sigma^+_j\sigma^-_k\sigma^-_l , \qquad \mbox{case (i)} ,
\end{equation}
where $\{\zeta_1,\zeta_2,\zeta_3,\zeta_4\}$ correspond to the four indexes $\{i,j,k,l\}$
ordered from the smallest to the largest one.
The remaining non-zero cases are those with (ii) $i=k, j=l$, $i=k, j\neq l$ and (iii) $i\neq k, j=l$
[the latter one being traceable back to case (ii)], which can be written as
\begin{eqnarray}
  c_i^\dagger c_j^\dagger c_i^{\phantom{\dagger}} c_j^{\phantom{\dagger}} & = &
  - \sigma^+_i\sigma^-_j\sigma^+_i\sigma^-_j ,  \hspace{1.95cm} \mbox{case (ii)} ,\: \\
  c_i^\dagger c_j^\dagger c_i^{\phantom{\dagger}} c_l^{\phantom{\dagger}} & = &
  \prod_{\zeta=\zeta_1+1}^{\zeta_2-1} \sigma^z_{\zeta}\,\, \sigma^+_i\sigma^+_j\sigma^-_i\sigma^-_l , \qquad \mbox{case (iii)} .
\end{eqnarray}
\end{subequations}

Conventional ED approaches to classically simulate the SYK Hamiltonian, although being highly inefficient, make use of such mapped nonlocal spin-$1/2$ network.
To overcome these limitations, proposals involving analog quantum simulations with ultracold gases~\cite{Danshita_2017, Wei_2021}, cavity QED~\cite{Uhrich_2023, Baumgartner_2024}, as well as
solid-state systems~\cite{Pikulin_2017, Chew_2017, Chen_2018, Brzezinska_2023} have been recently put forward.
Unfortunately, from an experimental point of view, it is very difficult to simulate strong both randomness and fully nonlocal interactions. Moreover even the initialization of the system into specific states at a given temperature and the measurement of the dynamical properties poses some conceptual issues.
For all these reasons, the most promising strategy seems to be the use of a digital quantum simulator~\cite{Solano_2017, Luo_2019, Babbush_2019} on a real quantum computer hardware, with individual and high-fidelity controllability. Viable platforms in this respect are superconducting qubits or nuclear spin systems.

\subsection{Failure of the quasiparticle picture}
\label{sec:SYK_quasiparticle}

Coming back to the reasons why the physics of the SYK model is so special in the quantum many-body scenario, we start from recalling the concept of {\it quasiparticle}, one of the cornerstones in modern
condensed matter physics, being crucial to describe a variety of situations in quantum matter.
For example, metals and semiconductors can be seen as free-Fermi systems, where each single electron
is surrounded by a cloud of other electrons, effectively renormalizing its characteristics;
this leads to the so called ``dressed fermion'' (i.e., quasiparticle) that behaves
exactly as a free electron, although with modified mass, magnetic moment, etc.
Other paradigmatic situations rely on analogous grounds, such as the theory of superconductivity
(pairing of quasiparticles), disordered metals and insulators (physics of localization), one-dimensional
materials (collective modes are seen as quasiparticles), fractional quantum Hall effect
(where quasiparticles are seen as ``fractions'' of an electron).
The SYK model lies outside this conventional scenario, since it cannot be described through the paradigm of quasiparticles.
For this reason, it is often referred to as a model for a strange metal~\cite{SYK_rev}.

At low energies, the physics of a weakly interacting fermionic system
is faithfully characterized by collective excitations that constitute
the normal modes of the associated Hamiltonian.
Let us suppose that $\{ \varepsilon_\alpha \}$ and $\{ n_\alpha \}$ are the eigenenergies
and the occupation numbers associated to quasiparticles of a given theory.
The energy of the system can be cast in the following simple form:
\begin{equation}
  E = \sum_\alpha n_\alpha \varepsilon_\alpha + \sum_{\alpha,\beta} n_\alpha n_\beta F_{\alpha \beta} + \ldots ,
  \label{eq:Ener_pert}
\end{equation}
where the coupling constants $F_{\alpha \beta}$ between two quasiparticles are small and can be treated
perturbatively, while further terms involving interactions between more than two particles are suppressed.
Systems undergoing this paradigm are characterized by a many-body energy spectrum that 
obeys some simple and universal rules: assuming there are $Q$ weakly interacting quasiparticles,
one can put forward a general scaling behavior for the gap $\Delta_0$ between the ground state
and the first excited state, as well as for the typical energy gap $\Delta$ between two generic
eigenstates in the middle of the spectrum.
In the thermodynamic limit, $\Delta_0$ is expected to close polynomially with $Q$,
while $\Delta$ should be exponentially suppressed with $Q$:
\begin{subequations}
  \begin{align}
    \Delta_0 \sim Q^{-r} \qquad & \mbox{with} \quad r = O(1); \label{eq:gap_poly} \\
    \Delta \sim e^{-\gamma Q} \qquad & \mbox{with} \quad \gamma \sim \log 2. \label{eq:gap_exp} 
  \end{align}
\end{subequations}
The above behaviors can be intuitively understood by counting the number of possible
combinations of quasiparticles (excitations), starting from their ground-state vacuum.
In fact, energies of quasiparticles are typically spaced polynomially with $Q$, while the total number
of many-body energy states should scale exponentially with $Q$. One thus expects a polynomial
spacing of low-lying quasiparticle excitations, in contrast with an exponential spacing
for excitations in the middle of the spectrum.

A simple model of a free-fermion theory in a disordered $N$-site lattice, falling into this paradigm,
is provided by the Hamiltonian
\begin{equation}
  H = \frac{1}{\sqrt{N}} \sum_{i,j} t_{ij} c^\dagger_i c_j ,
\end{equation}
where the prefactor $1/\sqrt{N}$ ensures extensivity and the hopping strength $t_{ij}$
is a Gaussian variable such that $\langle \! \langle t_{ij} \rangle \! \rangle = 0$
and $\langle \! \langle |t_{ij}|^2 \rangle \! \rangle = t^2$, analogously as for the couplings $J_{ijkl}$
in the SYK Hamiltonian~\eqref{eq:SYK_ham}.
Any given realization of the tunneling amplitudes $t_{ij}$ corresponds to different eigenenergies,
such that a diagonalization of the corresponding $N \times N$ hopping matrix maps the problem in noninteracting
normal modes [i.e., $F_{\alpha \beta} = 0$ in Eq.~\eqref{eq:Ener_pert}] which, of course, cannot thermalize.
Switching on a weak interaction term between such quasiparticles eventually leads to equilibration,
although with a long thermalization time $\tau_{\rm eq}$ (quasiparticles are still almost eigenstates of the system).
In fact, for a Fermi liquid close to the Fermi energy $E_F$, one finds the behavior
\begin{equation}
  \tau_{\rm eq} \sim \frac{\hbar E_F}{k_B^2 T^2}, \qquad \mbox{for} \quad T \to 0,
  \label{eq:tau_fermi}
\end{equation}
which is typical of normal metals (see, e.g., Refs.~\cite{Giuliani-Vignale_2005, Altland-Simons_2010}).
In the following, we will work in natural units, where the Boltzmann constant $k_B$, as well as the
reduced Planck constant $\hbar$, are set equal to one.

On the other hand, the SYK model for a strange metal violates the above paradigm.
In particular, it is no longer true that low-lying excitations are polynomially spaced
in energy [cf.~Eq.~\eqref{eq:gap_poly}], nor that the thermalization time diverges
as $\tau_{\rm eq} \sim T^{-2}$ [cf.~Eq.~\eqref{eq:tau_fermi}].
The reason eventually resides in the fact that two-body interactions dominate, so that
any perturbative treatment is impossible and the quasiparticle picture inevitably fails.
In principle, one could diagonalize the full Hamiltonian model in Eq.~\eqref{eq:SYK_ham},
thus obtaining $2^N$ normal modes (and not simply $N$ normal-mode quasiparticles).
In fact, numerical simulations for finite $N$ have shown that the gap
between the ground state and the first excited state at half filling is
\begin{equation}
  \Delta_0 \sim e^{-S_0 N}, \qquad \mbox{with} \quad S_0 \approx 0.465 .
  \label{eq:gap_delta0}
\end{equation}

\subsection{Large-$N$ limit}
\label{sec:SYK-LargeN}

Although finding the exact form of eigenvalues and eigenvectors of the SYK model is a hard task, a formally exact treatment can be derived in the limit of large $N$.
In particular, one can obtain closed analytic expressions for the fermionic Green's function
and for the free energy, which can then be processed numerically.
Following Refs.~\cite{Sachdev-Ye_1993, Fu-Sachdev_2016}, below we outline how to employ a path integral approach
on fermionic fields in imaginary time~\cite{Altland-Simons_2010}, to calculate the disorder-averaged
partition function and obtain the saddle-point equations for the Euclidean action
in the $N \to \infty$ limit.

\subsubsection{Euclidean action}

We use the formalism of coherent states.
Such states $\ket{\psi}$ are eigenstates of the fermionic annihilation operators $c_i$, such that
$c_i \ket{\psi} = \psi_i \ket{\psi}$ and $\bra{\psi} c_i^\dagger = \bra{\psi} \Bar{\psi}_i$.
In the fermionic case, the fields $\psi_i$ are anticommuting generators of a Grassmann algebra:
$\psi_i\psi_j = -\psi_j\psi_i$, where $\psi$ and $\Bar{\psi}$ have to be treated as independent variables.
A useful property of coherent states is that any normal ordered operator written the their basis
can be cast as a Grassmann number, by substituting $c_i \to \psi_i$ and $c^\dagger_j \to \Bar{\psi}_j$.
The partition function $Z$ of model~\eqref{eq:SYK_ham} can be thus written
as a field integral over such Grassmann variables:
\begin{equation}
  Z = {\rm Tr} \Big[ e^{-\beta( H_{\rm SYK} - \mu N)} \Big]
  = \int \mathcal{D}(\Bar{\psi},\psi) \, e^{-S[\Bar{\psi},\psi]} ,
\end{equation}
where $\beta = 1/T$ is the inverse temperature.
Here we use the replica trick ($n$ counts the number of replicas), introduce the Feynman differential
\begin{equation}
  \mathcal{D}(\Bar{\psi},\psi) \equiv \lim_{N\to \infty} \: \prod_{i=1}^N \prod_{a=1}^n \, \mathrm{d}\Bar{\psi}_i^{(a)}
  \mathrm{d}\psi_i^{(a)}
\end{equation}
and the imaginary-time ($\tau$) Euclidean action
\begin{equation}
  S[\Bar{\psi},\psi] = \!\int_{0}^{\beta} \!\mathrm{d} \tau\, \sum_a \bigg\{\sum_i \Bar{\psi}_i^{(a)}(\tau)\left(\frac{\partial}{\partial\tau} \!-\! \mu\right)\psi_i^{(a)}(\tau)
  + \!\! \sum_{i,j,k,l=1}^N \!\! J_{ijkl}\left.\Bar{\psi}_i^{(a)} \Bar{\psi}_j^{(a)} \psi_k^{(a)} \psi_l^{(a)} \right\vert_\tau \bigg\} .
\end{equation}

The action and the partition function depend on the specific realization of the coupling constants $J_{ijkl}$,
it is thus useful to consider averages over the disorder in such a way to obtain the following
modified average partition function
\begin{eqnarray}
    \langle \! \langle Z \rangle \! \rangle & = & \int \mathrm{d}(J_{ijkl}) \,\rho(\{J_{ijkl}\})\, \mathcal{Z}(\{J_{ijkl}\})
   \nonumber \\
    & = & \mathcal{N}\int \mathrm{d}(J_{ijkl}) \,\exp\left\{-\frac{N^3}{2J^2} \,\sum_{i,j,k,l} J_{ijkl}^2\right\}\, \int \mathcal{D}(\Bar{\psi},\psi)\,e^{-S\left[\Bar{\psi},\psi, \{J_{ijkl}\}\right]} \nonumber\\
    & = & \mathcal{N}\int \mathrm{d}(J_{ijkl}) \int \mathcal{D}(\Bar{\psi},\psi)\, e^{-\Tilde{S}\left[\Bar{\psi},\psi, \{J_{ijkl}\}\right]},
\end{eqnarray}
where $\mathcal{N}$ is an irrelevant normalization constant
and the modified Euclidean action $\Tilde{S}$ absorbs the functional form of the Gaussian distribution for
the $J_{ijkl}$'s.
Completing the square in the action, we can write
\begin{eqnarray}
  \Tilde{S} & = & \sum_{i,a} \int_{0}^{\beta} \mathrm{d}\tau\,\Bar{\psi}_i^{(a)}(\tau)
  \left(\frac{\partial}{\partial\tau} - \mu\right) \psi_i^{(a)}(\tau) \nonumber \\
  && + \frac{1}{2}\sum_{i,j,k,l} \bigg\{ \frac{N^{3/2}J_{ijkl}}{J} + \frac{J}{N^{3/2}} \sum_a \int_0^{\beta}
  \mathrm{d}\tau\,\left.\Bar{\psi}_i^{(a)} \Bar{\psi}_j^{(a)} \psi_k^{(a)} \psi_l^{(a)} \right\vert_\tau \bigg\}^2 \nonumber \\
  && - \frac{J^2}{2N^3} \sum_{i,j,k,l} \, \sum_{a,b} \, \iint_0^\beta \mathrm{d}\tau \, \mathrm{d}\tau' \,
  \left.\Bar{\psi}_i^{(a)} \Bar{\psi}_j^{(a)} \psi_k ^{(a)}\psi_l ^{(a)} \right\vert_\tau\left.
  \Bar{\psi}_i^{(b)} \Bar{\psi}_j^{(b)} \psi_k^{(b)} \psi_l^{(b)} \right\vert_{\tau'} .
  \label{eq:Modif_act}
\end{eqnarray}
Now, performing the Gaussian integral over the coupling constants $\{J_{ijkl}\}$
one reabsorbes the second line of Eq.~\eqref{eq:Modif_act} in the measure of the functional integral,
as well as washes out the normalization constant $\mathcal{N}$.
Moreover, in the large-$N$ limit, the normal-ordering corrections in the third line can be neglected, to finally obtain
\begin{equation}
  \Tilde{S} = \sum_{i,a} \int_{0}^{\beta} \mathrm{d}\tau\,\Bar{\psi}_i^{(a)}(\tau) \left(\frac{\partial}{\partial\tau}
  - \mu\right)\psi_i^{(a)}(\tau) - \frac{J^2}{2N^3} \, \sum_{a,b} \iint_0^\beta \mathrm{d}\tau\,\mathrm{d}\tau'\,
  \bigg\vert \sum_i\Bar{\psi_i}^{(a)}(\tau) \psi_i^{(b)}(\tau') \bigg\vert^4.
\end{equation}

\subsubsection{Hubbard-Stratonovich transformation}

To decouple the interaction term containing eight Grassmann fields, one can perform two subsequent
Hubbard-Stratonovich transformations. This can be done by introducing two auxiliary scalar fields $G$ and $\Sigma$,
which couple to the Grassmann fields, such that the residual interaction is quadratic in the fields themselves.
We only consider the diagonal solution in the replica space and omit the corresponding replica index.
It is thus possible to write
\begin{equation}
     \langle \! \langle Z \rangle \! \rangle  = \iint \mathrm{d} G\,\mathrm{d}\Sigma \int \mathcal{D}(\Bar{\psi},\psi) \,e^{-S\left[\Bar{\psi},\psi, G,\Sigma\right]} ,
\end{equation}
with
\begin{eqnarray}
  S\left[\Bar{\psi},\psi, G,\Sigma\right] & = & \sum_i \int_{0}^{\beta} \mathrm{d}\tau\,\Bar{\psi}_i(\tau)
  \left(\frac{\partial}{\partial\tau} - \mu\right) \psi_i(\tau) \nonumber \\
  && + \iint_0^\beta \mathrm{d}\tau\,\mathrm{d}\tau'\,\Sigma(\tau,\tau') \left[N G(\tau,\tau')-\sum_i\Bar{\psi}_i(\tau)\psi_i(\tau')\right] \nonumber \\
  && + \frac{NJ^2}{2} \iint_0^\beta \mathrm{d}\tau\,\mathrm{d}\tau'\, G^2(\tau,\tau') \, {G^*}^2(\tau,\tau') .
\end{eqnarray}

If one now writes the saddle-point equations $\frac{\delta S}{\delta \Sigma} = \frac{\delta S}{\delta G^*} = 0$,
it is possible to recognize that $G$ coincides with the imaginary-time-ordered retarded Green's function
\begin{equation}
    G(\tau,\tau') \equiv - \left\langle \mathcal{T}_\tau\left[ c(\tau) \, c^\dagger(\tau')\right]\right\rangle.
\end{equation}
Assuming time-translation invariance and defining the fermionic Matsubara frequencies
$\omega_n = (2n+1)\pi/\beta$, ($n \in \mathbb{Z}$), the saddle-point equations finally read:
\begin{subequations}
  \label{eqs:largeN_syk}
  \begin{eqnarray}
    G(i\omega_n) & = & \frac{1}{i\omega_n +\mu - \Sigma(i\omega_n)} , \\
    \Sigma(\tau) & = & - J^2 G^2(\tau) \, G(-\tau) .
  \end{eqnarray}
\end{subequations}
Although these are algebraic equations, rigorously exact in the thermodynamic limit, their solution
is neither obvious nor immediate to be found. In fact they are connected non locally through
an imaginary-time Fourier transform and they need to be solved numerically.
As reported in Sec.~\ref{sec:Green}, their solution gives a behavior
${\rm Im} \{ G(\omega) \} \sim \omega^{-1/2}$ for $\omega \to 0$,
in stark contrast with a normal Fermi-liquid behavior.

\subsection{Main properties of the SYK model}

We are now in the position to summarize some important features of the SYK Hamiltonian, emphasizing the properties that lead it to a prototypical nonperturbative model of strongly correlated fermions.
All the results discussed below have been obtained either through the large-$N$ limit solution, or by classical ED simulations for small system sizes. Possible forthcoming quantum simulations could pave the way for a further and deeper understanding of the physics of this model.

\subsubsection{Particle-hole symmetry}

First of all, the Hamiltonian defined in Eq.~\eqref{eq:SYK_ham} is not symmetric under particle-hole (PH) exchange.
In fact, given the operator 
\begin{equation}
  \mathcal{P} \equiv \prod_i \bigl( c_i^\dagger + c_i \bigr) \, \mathcal{K},
  \qquad \text{with} \quad \mathcal{K} \mathcal{O} = \mathcal{O}^* ,
\end{equation}
which exchanges particles into holes and vice versa, one finds $[ \mathcal{P}, H_{\text{SYK}} ] \neq 0$.
The reason comes from terms with two equal indexes.
It is however possible to construct a modified SYK model which exploits the PH symmetry, by explicitly
subtracting terms in Eq.~\eqref{eq:SYK_ham} associated with two equal indexes:
\begin{equation}
  H_{\text{SYK}}^{\text{(PH)}} = H_{\text{SYK}} + \frac{1}{(2N)^{3/2}} \sum_{i,j,k,l}^N \frac{\tilde J_{ijkl}}{2}
  \left( \delta_{ik} c^\dagger_jc_l - \delta_{il} c^\dagger_jc_k - \delta_{jk} c^\dagger_i c_l + \delta_{jl} c^\dagger_i c_k \right)
\end{equation}
We point out that, in any case, in the thermodynamic limit $N\to\infty$,
contributions with two equal indexes are suppressed as $\mathcal{O}(N^{-1})$,
thus the original model restores the PH symmetry:
$H_{\text{SYK}}^{\text{(PH)}} = H_{\text{SYK}}$ for $N \to \infty$.

\subsubsection{Entropy}

\begin{figure}
  \begin{centering}
  \includegraphics[width=0.99\columnwidth]{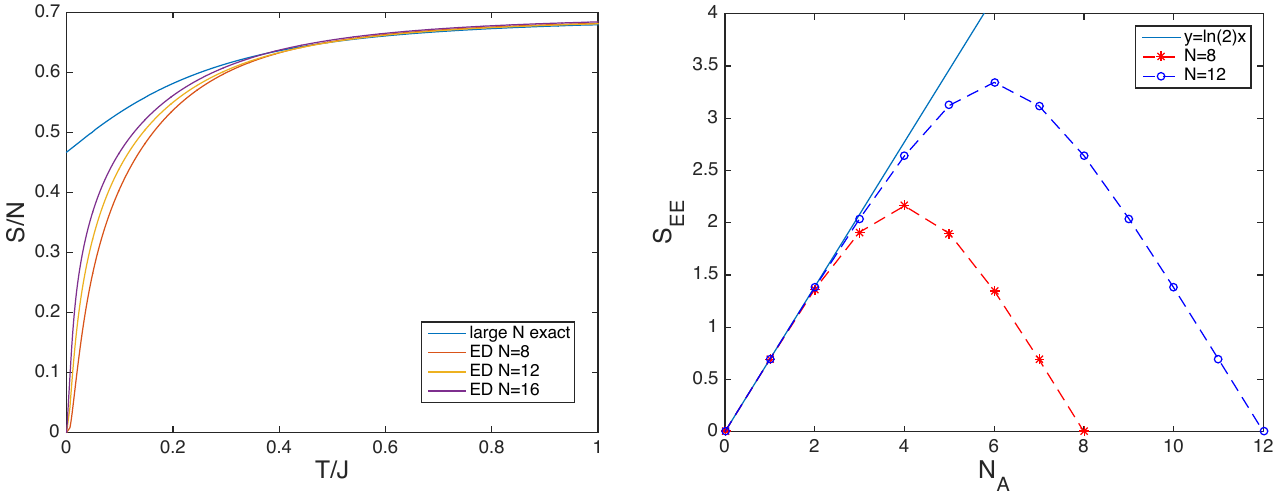}
  \caption{Some results for the entropy behavior in the fermionic SYK model~\eqref{eq:SYK_ham},
    obtained either from ED results (with $N$ up to 16) or semi-analytically in the large-$N$ limit.
    (left) The thermal entropy as a function of the temperature.
    (right) The zero-temperature bipartite entanglement entropy as a function of the subsystem size $N_A$,
    for a given system size $N$.
    Adapted from Ref.~\cite{Fu-Sachdev_2016}.}
  \label{fig:SYK_S}
  \end{centering}
\end{figure}

The finite-temperature thermodynamic entropy can be obtained from the free energy $F$.
In fact, according to the rules of thermodynamics, one has:
\begin{equation}
  \frac{S}{N} = -\frac{1}{N} \frac{\partial F}{\partial T},  
  \qquad \mbox{with} \quad \frac{F}{N} = - \frac{\log Z}{N \beta}.
\end{equation}
The partition function $Z = \sum_n e^{-\beta E_n}$ can be obtained either by explicitly computing
the full Hamiltonian spectrum through exact-diagonalization (ED) methods
(details are provided in Sec.~\ref{subsec:SYK_simul}),
or in a semi-analytic way, from the modified action approach (see Sec.~\ref{sec:SYK-LargeN}).
In fact, one can show that~\cite{Fu-Sachdev_2016}
\begin{equation}
  \frac{F}{N} = \frac{1}{\beta} \sum_n \log \big[ -\beta G(i \omega_n) \big]
  - \frac{3}{4 \beta} \sum_n \Sigma(i \omega_n) \, G(i \omega_n). 
\end{equation}
Numerical data displayed in Fig.~\ref{fig:SYK_S}(left) show that, at high temperatures, the entropy per site tends
to the maximum admissible value $S_\infty/N = \ln 2 \approx 0.693$, while, in the zero-temperature limit
and for finite system sizes, it approaches the finite value $S_0/N \approx 0.465$.
The latter corresponds to the rate at which the gap between the lowest energy levels closes
[see Eq.~\eqref{eq:gap_delta0}].
Interestingly, at low temperatures, the ED results at finite $N$ show a marked discrepancy with the exact
solution in the large-$N$ limit: this is due to the non commutativity of the thermodynamic limit ($N \to \infty$)
with the low-temperature ($T \to 0$). In fact, the non-zero entropy
for $T \to 0$ can be obtained only after taking $N \to \infty$ first~\cite{Fu-Sachdev_2016}.

%%%%%%%%%%%%%%%%%%%%%%%%%%%%%%%%%%%%%%%%%%%%%%%%%%%%%%%%%%%%%%%%%%%%%%%%%%%%%%%%%%%%%%%%%%%%%%%%%%%%%%%%
%%%%%%%%%%%%%%%%%%%%%%%%%%%%%%%%%%%%%%%%%%%%%%%%%%%%%%%%%%%%%%%%%%%%%%%%%%%%%%%%%%%%%%%%%%%%%%%%%%%%%%%%

One can also compute the ground-state entanglement entropy for a bipartition $A|B$ of the whole system $A+B$,
through the von Neumann entropy of the reduced density matrix $\rho_A$ of subsystem $A$:
\begin{equation}
  S_{EE} = - {\rm Tr} \big[ \rho_A \ln \rho_A \big], 
\end{equation}
where $\rho_A = {\rm Tr}_B \big[ |0\rangle \, \langle 0| \big]$ is obtained by tracing over the sites
corresponding to the complementary subsystem $B$. Here $|0\rangle$ denotes the system global ground state.
Let us suppose that subsystem $A$ contains $N_A < N$ sites.
Numerical results displayed in Fig.~\ref{fig:SYK_S}(right) show that, for $N_A < N/2$, the entanglement entropy
is proportional to $N_A$, thus obeying a volume-law behavior. Therefore one expects thermalization to
occur even in the ground state. This is a remarkable feature of the SYK model, in sharp contrast
with typical low-dimensional systems where an emerging area-law behavior is expected at
low energies~\cite{Eisert_2010}.
It is worth pointing out that, while the ratio $S_{EE}/N_{A}$ should correspond to the zero-temperature
limit of the entropy per site $S_0/N$, numerical results appear to be closer to the opposite
infinite-temperature limit $S_\infty/N$. Such qualitative discrepancy has been ascribed
to finite-size effects~\cite{Fu-Sachdev_2016}.

\subsubsection{Green's function}
\label{sec:Green}

The on-site retarded Green's function
\begin{equation}
    G^R_i(t,t') = - i \Theta(t-t') \left\langle \left\{ c_i^\dagger(t), c_i(t')\right\} \right\rangle
\end{equation}
can be faithfully studied within the Lehmann representation:
\begin{equation}
  G^R_i(\omega) = \frac{1}{Z} \sum_{n,m} \frac{ \langle n| c_i | m \rangle
    \langle m | c_i^\dagger | n \rangle}{\omega + E_n - E_m + i\eta} \left(e^{-\beta E_n} + e^{-\beta E_m}\right) ,
\end{equation}
where $\{ |n\rangle \}$ are the eigenstates of $H_{\rm SYK}$ corresponding to the eigenenergies $\{ E_n \}$
and $\eta > 0$ is a small regularization parameter.
In the zero temperature limit, we can cast the latter expression as
\begin{equation}
  G^R_i(\omega) =  \sum_{m} \frac{ \langle 0| c_i | m \rangle
    \langle m | c_i^\dagger | 0 \rangle}{\omega + E_0 - E_m + i\eta} + \frac{ \langle 0| c_i | m \rangle
    \langle m | c_i^\dagger | 0 \rangle}{\omega - E_0 + E_m + i\eta} .
  \label{eq:Green_num1}
\end{equation}

%%%%%%%%%%%%%%%%%%%%%%%%%%%%%%%%%%%%%%%%%%%%%%%%%%%%%%%%%%%%%%%%%%%%%%%%%%%%%%%%%%%%%%%%%%%%%%%%%%%%%%%%
\begin{figure}
  \begin{centering}
  \includegraphics[width=0.8\columnwidth]{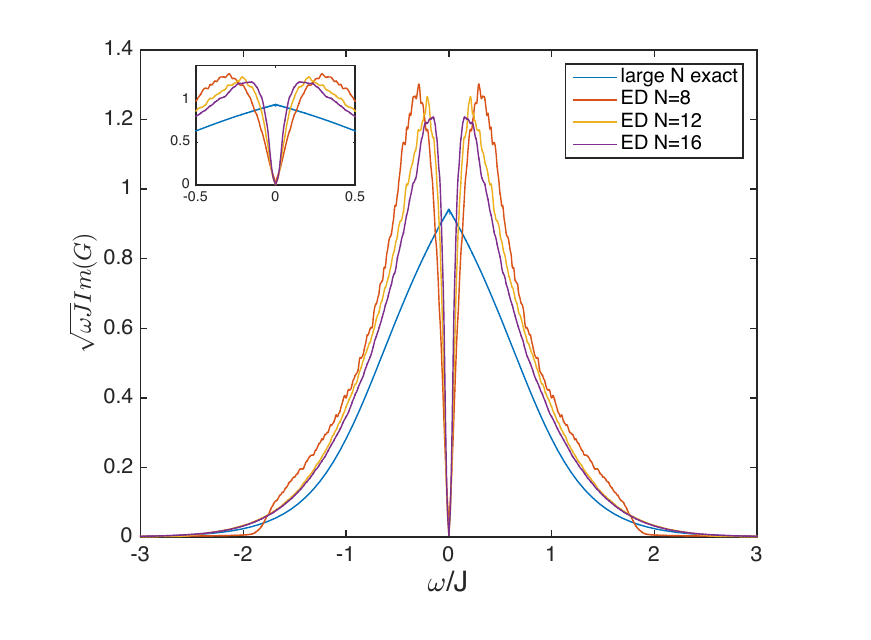}
  \caption{Behavior in frequency space of the imaginary part of the retarded Green's function for the SYK model~\eqref{eq:SYK_ham}.
    the inset is a zoom close to $\omega = 0$.
    From Ref.~\cite{Fu-Sachdev_2016}.}
  \label{fig:SYK_Green}
  \end{centering}
\end{figure}
%%%%%%%%%%%%%%%%%%%%%%%%%%%%%%%%%%%%%%%%%%%%%%%%%%%%%%%%%%%%%%%%%%%%%%%%%%%%%%%%%%%%%%%%%%%%%%%%%%%%%%%%

It is now useful to compare the results for the imaginary part of the Green's function $G(\omega)$
in the large-$N$ limit, which can be obtained from Eqs.~\eqref{eqs:largeN_syk},
with those at finite $N$, numerically computable from Eq.~\eqref{eq:Green_num1},
given the fact that $G^R_i(\omega) = G(i \omega_n \to \omega + i \eta)$.
Using the fact that $(\omega + i \eta)^{-1} = \mathcal{P}\tfrac{1}{\omega} - i \pi \delta(\omega)$,
one can see that the imaginary part of $G^R_i(\omega)$ contains a Dirac delta,
which can be numerically estimated through a Lorentzian, using the well known limit
\begin{equation}
  \delta\left(E_0 -E_m +\omega\right) = \lim_{\eta \to 0^+} \frac{1}{\pi} \frac{\eta}{(E_0-E_m+\omega)^2+\eta^2}
\end{equation}
We also recall that, for $\mu=0$, one has to keep into account the anti-unitary PH symmetry, which gives rise
to double degenerate ground states for certain system sizes. In such cases, one needs to sum over both
ground states $|0\rangle$ and $|0'\rangle$, in the expression for the propagator.
Figure~\ref{fig:SYK_Green} compares the imaginary part of the retarded Green's function as obtained
either in the large-$N$ limit (blue curve) or at finite size (other colored curves)~\cite{Fu-Sachdev_2016}.
In the latter cases, ED results do not faithfully reproduce the peculiar $\sim \omega^{-1/2}$ divergence
at low frequencies, typical of strange metals (to be contrasted with the usual $\sim \omega^{-2}$ behavior
for normal metals), which is clearly seen in the thermodynamic limit.

\subsubsection{Quantum chaos and scrambling}
\label{scrambling}
 
A prominent feature of the SYK model is that it exhibits a quantum chaotic behavior~\cite{Maldacena-etal_2016}.
To better highlight its meaning, we start from the heuristic definition of {\em chaos} in the classical realm.
Let us consider a classical dynamical system and the corresponding phase space $\Gamma$,
in which we can define a notion of measure.
Chaos sets in whenever, for two generic initial conditions $x_1(0), x_2(0) \in \Gamma$ close together,
the distance $\delta x(t) \equiv \vert x_1(t) - x_2(t) \vert$ between the two originating trajectories
in the phase space grows exponentially in time: $\delta x(t) \sim \delta x(0) e^{\lambda_L t}$,
where $\lambda_L > 0$ is the largest classical Lyapunov exponent of the classical system.
More precisely, the latter is defined as
\begin{equation}
  \lambda_L = \lim_{t \to \infty} \, \lim_{\delta x(0) \to 0} \, \frac{1}{t} \ln \frac{\delta x(t)}{\delta x(0)} .
\end{equation}

To generalize this notion to quantum systems, is is useful to rewrite the above sensitivity
in terms of a Poisson bracket:
\begin{equation}
  \frac{\delta x(t)}{\delta x(0)} = \big\{ x(t), p(0) \big\}_{\text{cl}} ,
\end{equation}
where $p(0)$ is the initial momentum. We can then proceed with the canonical quantization prescription,
promoting the variables $x$ and $p$ to operators and substituting the Poisson brackets with the commutator:
$ \{ \cdot\,, \cdot \}_{\text{cl}} \to -i [ \, \cdot , \cdot \, ]$.
When performing expectation values of this kind of observable, cancellations may set in,
due to terms with opposite sign. In general, it is thus better to take the norm squared of this operator:
\begin{equation}
  \Big\{ - i \big[ x(t), p(0) \big] \Big\} \, \Big\{ i \big[ x(t),p(0) \big]^\dagger \Big\}
  = - \big[ x(t) , p(0) \big]^2 .
\end{equation}
This is an example of an out-of-time-order correlator (OTOC), whose expectation value is believed
to provide insight and characterize quantum chaos~\cite{Larkin_1969, Cotler_2018}.

In a chaotic quantum system at thermal equilibrium and at early times,
the expectation value of the above OTOC is expected to grow exponentially as 
\begin{equation}
  \left\langle[x(t), p(0)]^2 \right\rangle \sim c_0 + c_1 e^{\lambda_Lt} .
\end{equation}
The time scale at which such correlator becomes significantly is often referred to
as the ``scrambling time", since it roughly defines the onset of
quantum information spreading (or ``scrambling") over the system.
Furthermore it can be shown that there is an upper thermal limit for this growth,
provided by $\lambda_L \lesssim 2\pi T$~\cite{Maldacena-etal_2016}, which is expected to saturate
in the strong-coupling limit ($\beta J \gg 1$) of the SYK model~\cite{Kitaev_2015}, as verified
by extensive numerical simulations~\cite{Fu-Sachdev_2016, Roberts_2018, Kobrin_2021}.
Such Lyapunov exponent represents the equilibration rate for the system, for which
one can define a thermalization time scale behaving as
\begin{equation}
    \tau_{\text{eq}} \sim T^{-1} .
\end{equation}
This result qualitatively differs from the normal metal behavior of Eq.~\eqref{eq:tau_fermi}, 
which presents a much slower equilibration $\propto T^{-2}$.

Explicit numerical simulations of the time behavior for the OTOC $\langle x(t) p(0) x(t) p(0) \rangle$ at infinite temperature evidence an exponential decay in time of such correlator, which translates into an exponential growth of $\left\langle[x(t), p(0)]^2 \right\rangle$~\cite{Fu-Sachdev_2016, Roberts_2018, Kobrin_2021}.

\section{Case study of a complex quantum dynamics}
\label{usefulSYK}

At this stage, we are ready to discuss a physical example where the peculiar dynamic properties of a complex quantum network emerge and 
can be exploited for quantum technological purposes.
In particular, we refer to the possibility to build up a concept of quantum battery (QB)~\cite{Quack_2023}
that could take advantage of the fast thermalization time scales and scrambling properties
achieved by the dynamics of a complex quantum network over more traditional quantum materials.
The goal is to outperform conventional classical schemes, where the charging efficiency
of the battery typically scales linearly with the number of its components.

\subsection{The quantum battery paradigm}

The concept of QB has been put forward a dozen of years ago~\cite{Alicki_2013, Hovhannisyan_2013},
in the context of work extraction from quantum systems that are used to temporarily store energy~\cite{Allahverdyan_2004}.
In its basic formulation, a quantum battery (QB) can be identified by a quantum system
that is capable to accumulate energy, when undergoing a double quantum quench~\cite{Campaioli-rev_2024}.
Specifically, the typical charging protocol of a QB performs as follows.
We consider a system described by a time-independent Hamiltonian
\begin{equation}
  H_0 = \sum_{j=1}^N h_j ,
  \label{eq:QB_ham0}
\end{equation}
where $N$ denotes the number of quantum cells, each of them being initially decoupled
from the rest of the system and governed by a local Hamiltonian term $h_j$.
In the hypothesis that the global system is in a generic (mixed) state $\rho_N(t)$,
the total internal energy can be calculated as
\begin{equation}
  E_N(t) = {\rm Tr} \big[ \rho_N(t) \, H_0 \big] = \sum_{j=1}^N {\rm Tr} \big[ \rho_N(t) \, h_j \big].
  \label{eq:QB_energy}
\end{equation}
At a given reference time, say $t=0$, the Hamiltonian is instantaneously changed
into $H_I$, which includes interactions (I) between the various quantum cells,
so that the system evolves unitarily under the new Hamiltonian.
Subsequently, after a given time $\tau$ (the so-called charging time), a second quench
is performed and the system comes back to be described by the original Hamiltonian $H_0$
in Eq.~\eqref{eq:QB_ham0}.
Thus, the Hamiltonian governing the system along the whole protocol can be cast in the form
\begin{equation}
  H(t) = \bigl[ 1-\lambda(t) \bigr] H_0 + \lambda(t) H_I
  \label{eq:QB_ham}
\end{equation}
where $\lambda(t)$ is a double step function which can take either the value 1 for $0<t<\tau$,
or be zero otherwise.
For the sake of simplicity, in the following we neglect any dissipative mechanism
and suppose that the QB is initialized in the (nondegenerate) ground state $|\psi_N(0) \rangle$ of $H_0$,
so that, under the above protocol, the system remains in a pure state.

Let us now introduce the quantities in which one might be interested.
Denoting with $|\psi_N(t) \rangle$ the evolved state of the system at time $t$,
one finds:
\begin{equation}
  |\psi_N(t) \rangle = e^{- i H_I t} |\psi_N(0)\rangle.
\end{equation}
It is clear that the energy of the QB is constant at any time, with the exception
of the quenching points $t=0$ and $t=\tau$, and is given by Eq.~\eqref{eq:QB_energy}
with $\rho_N(t) = |\psi_N(t) \rangle \, \langle \psi_N(t)|$.
The total energy $E_N(\tau)$ injected into the $N$ quantum cells at the end of the protocol
can be thus expressed in terms of the mean local energy.
It is also possible to define the corresponding power of the QB as $P_N(\tau) = E_N(\tau) / \tau$.
To ensure a non trivial unitary evolution in the time $t \in [0,\tau]$
between the two quenches (and thus to effectively charge the system)
one must require that $[ H_0, H_I ] \neq 0$.

It is worth mentioning that the energy fraction which can be effectively extracted
for thermodynamic purposes, when acting unitarily on $M \leq N$ quantum cells
and without having access to the full system, does not necessarily coincide with the total energy $E_N(\tau)$.
Indeed, part of the energy injected into the QB is in the form of correlations between the quantum cells
that develop in time, and is thus useless for practical purposes.
A rigorous measure of such useful energy fraction is provided by
the so-called ``ergotropy"~\cite{Allahverdyan_2004}
of the local state $\rho_M (t)$ of $M$ cells, defined as the difference 
\begin{equation}
  \mathcal{E}_M(t) = E[\rho_M(t)] - E[\tilde \rho_M(t)]
\end{equation}
between the energy $E[\rho_M(t)] = {\rm Tr} [\rho_M(t) \, H_M]$ of the state $\rho$,
where $H_M = \sum_{j=1}^M h_j$ is the local Hamiltonian of $M$ cells,
and that of the passive counterpart $\tilde \rho_M(t)$ of $\rho_M(t)$, on the same local Hamiltonian.
The latter state $\tilde \rho_M(t)$ is defined as a density matrix which is diagonal
on the eigenbasis of $\mathcal{H}_M$ and whose eigenvalues correspond to a proper reordering
of those of $\rho_M$,
\begin{equation}
  \tilde \rho_M = \sum_n r_n |\epsilon_n\rangle \, \langle \epsilon_n|, \quad \mbox{where }
  \begin{array}{rll} \rho_M = & \!\!\! \sum_n r_n |r_n\rangle \, \langle r_n|, & r_0 \geq r_1 \geq r_2 \geq \cdots ,
    \vspace*{1mm}\\
    H_M = & \!\!\! \sum_n \epsilon_n |\epsilon_n\rangle \, \langle \epsilon_n|, &
    \epsilon_0 \leq \epsilon_1 \leq \epsilon_2 \leq \cdots ,
  \end{array}
\end{equation}
so that $E[\tilde \rho_M] = \sum_n r_n \epsilon_n$.
Note that, if $\epsilon_0 = 0$ and the state $\rho_M$ is pure, then $E(\tilde \rho_M) = 0$
and thus the ergotropy coincides with the energy of $\rho_M$.
On the other hand, if the state $\rho_M$ is mixed, one generally finds $\mathcal{E}_M(t) < E[\rho_M(t)]$
and thus the extractable work is smaller than the actual energy of $\rho_M$.

\subsection{The Dicke quantum battery}

We first consider the probably most iconic example of a QB, provided by the Dicke model, which describes an array of $N$ two-level atoms
inside a resonant cavity~\cite{Ferraro_2018}. For simplicity, we focus on the case in which the laser
in the cavity is resonant with the relevant atomic transition of the atoms (zero-detuning condition);
we also ignore all the other modes in the cavity.
Here the atoms will play the role of the quantum cells and their interaction
with the light will constitute the charging mechanism of the battery.
The system of interest is described by a time-dependent Hamiltonian of the same form as in
Eq.~\eqref{eq:QB_ham}:
\begin{equation}
  H(t) = \omega \Big[ a^\dagger a + J^z + 2 \lambda(t)(J^+ + J^-)( a^\dagger + a ) \Big]
  \label{eq:Dicke}
\end{equation}
where $a^{(\dagger)}$ are the canonical annihilation (creation) operators for photons in the cavity,
while $J^\alpha = \tfrac{1}{2}\sum_j \sigma^\alpha_j$
(with $\alpha=x,y,z,+,-$) is the total spin of the atomic component.
The first term in Eq.~\eqref{eq:Dicke} is the energy of the resonating mode,
the second one denotes the energy of the atoms, while the third one quantifies the dipolar
light-matter interaction.
In the rotating wave approximation, valid for a weak light-matter coupling,
all the interaction terms that do not conserve the number of excitations
(as $J^+ a$ and $J a^\dagger$) can be neglected. In that case, during the whole dynamics,
the system maintains the same number excitations of the initial state
and the Dicke model reduces to the well known, and much easier to solve, Jaynes-Cummings model.

Before discussing the proper QB Dicke model, as described by Eq.~\eqref{eq:Dicke}, we focus
on a simplified scenario in which, instead of taking a single cavity with all the atoms inside,
one puts each of the $N$ atoms in a separate cavity. All the cavities are treated independently,
so that one realizes a so-called ``parallel charging" (${}^\parallel$) protocol.
In this situation, the $j$-th atom ($j = 1, \ldots, N$) is described through the reduced local Hamiltonian
\begin{equation}
  h_j = \omega \Big[ a_j^\dagger a_j + \tfrac{1}{2} \sigma^z_j
    + \lambda(t)(\sigma^+_j + \sigma_j^-)( a_j^\dagger + a_j ) \Big].
\end{equation}
The total energy of $N$ atoms in $N$ cavities is thus equal to $N$ times the one of a single atom
and its corresponding cavity, so that $E^\parallel_N(\tau) \propto N$.
Likewise, the power is linear in the system size: $P^\parallel_N(\tau) \propto N$.

Coming back to the QB Dicke model of Eq.~\eqref{eq:Dicke},
photons inside the only cavity are able to mediate interactions between the $N$ atoms,
so that one realized a so-called ``collective charging" (${}^\perp$) protocol.
Starting from an initial factorized state of the form $|\psi_N(0)\rangle = | g, \dots ,g \rangle_N \otimes |N\rangle$,
where $|g\rangle$ denotes the ground state of one atom and $|N\rangle$ is the Fock
state of $N$ photons in the cavity, one finds again an extensive energy $E^\perp_N(\tau) \propto N$,
but the scaling of the power is superextensive: $P^\perp_N(\tau) \propto N^{3/2}$.
A more accurate analysis shows that the latter result is an artifact due to the improper
dependence of the light-matter coupling term in Eq.~\eqref{eq:Dicke} with the system size.
In fact, to ensure extensivity of the system, one should
rescale the coupling constant $\lambda(t)$ by a $1/\sqrt{N}$ factor, so that the interaction
strength decreases with the number of atoms. In such case, the power $P^\perp_N(\tau)$ would again
scale linearly with $N$, so that the collective charging would behave analogously as the parallel charging,
apart from a prefactor independent of the size~\cite{Lewenstein_2020}.

Several other QB toy models have been proposed and tested, using different interaction Hamiltonians to inject energy into the quantum cells (see, e.g., the review~\cite{Campaioli-rev_2024}). Unfortunately, the typical local interactions of
quantum many-body systems do not provide a charging speedup that is genuinely induced by the entanglement
developed along the dynamics; on the other hand, this is rather given by the lack
of a well defined thermodynamic limit, as for the ill-defined Dicke QB.
The SYK interaction falls beyond this paradigm and presents qualitative differences, as we discuss below~\cite{Rossini_2020, Rosa_2020}.

\subsection{The SYK model as a quantum battery}

The key point to promote the SYK model to a toy system realizing a conceptual QB
that outperforms more conventional schemes resides in adopting
its random two-particle interaction term [cf.~Eq.~\eqref{eq:SYK_ham}] as the charging Hamiltonian:
\begin{equation}
  H_I = \frac{1}{(2 N)^{3/2}} \sum_{i,j,k,l=1}^N \tilde J_{ijkl} \, c_i^\dagger c_j^\dagger c_k c_l .
\end{equation}
After performing the mapping of Sec.~\ref{subsec:SYK_simul} from the original fermionic formulation
into a spin-$1/2$ network, one can identify each cell of the QB with a single mapped spin~\cite{Rossini_2020}.
The battery, when not in charge, should be described by a time-independent spin Hamiltonian
which does not commute with $\mathcal{H}_I$. By convention, we consider
\begin{equation}
  H_0 = \omega J^y = \frac{\omega}{2} \sum_{j=1}^N \sigma_j^y,
\end{equation}
so that $[ H_0, H_I ] \neq 0$.
As for the initial state, we take the ground state of $H_0$,
$|\psi_N(0) \rangle = \bigotimes_{j=1}^N | \! \leftarrow \rangle_j$,
with $\sigma_j^y | \! \leftarrow \rangle_j = -| \! \leftarrow \rangle_j$. 
Extensive numerical calculations with up to $N = 16$ spins have shown that a QB based on such SYK charging scheme
displays a genuine superextensive scaling of its power output with $N$~\cite{Rossini_2020}.
This property is directly associated with the chaoticity of the model and its extremely fast thermalization time.

%%%%%%%%%%%%%%%%%%%%%%%%%%%%%%%%%%%%%%%%%%%%%%%%%%%%%%%%%%%%%%%%%%%%%%%%%%%%%%%%%%%%%%%%%%%%%%%%%%%%%%%%
\begin{figure}[!t]
  \begin{centering}
  \includegraphics[width=1.\columnwidth]{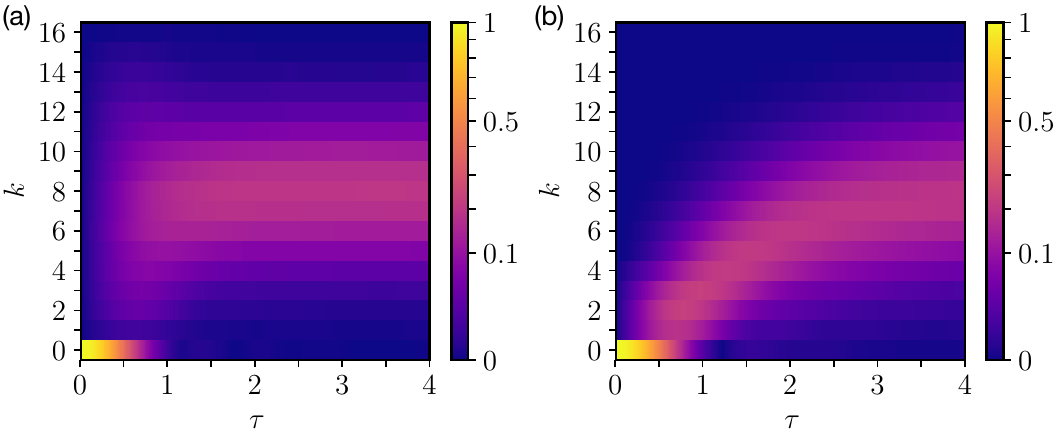}
  \caption{The population $p_k$ of each energy level of the QB as a function of the index $k$ and of the
    charging time $\tau$. Panel (a) is for a fermionic-SYK charging scheme, while panel (b)
    shows the same result for the bosonic-SYK counterpart. The reported data are for QBs with $N=16$ cells
    and correspond to a single disorder realization in the couplings.
    Adapted from Ref.~\cite{Rossini_2020}.}
  \label{fig:SYK_QB}
  \end{centering}
\end{figure}
%%%%%%%%%%%%%%%%%%%%%%%%%%%%%%%%%%%%%%%%%%%%%%%%%%%%%%%%%%%%%%%%%%%%%%%%%%%%%%%%%%%%%%%%%%%%%%%%%%%%%%%%

Focusing on the spectrum of $H_0$, one has $N+1$ equally spaced energy levels,
with a degeneracy associated to the $k$-th level that is equal to $g_k = \binom{N}{k}$.
The spectral decomposition of the Hamiltonian can be cast as a simple summation over the $k$ levels
and their degeneracy index $i$, which labels $g_k$-degenerate states:
\begin{equation}
  H_0 = \sum_k \varepsilon_k \sum_i |k,i \rangle \langle k,i| .
\end{equation}
The populations of the various states along the charging process are expressed as
\begin{equation}
  p_k(\tau) = \sum_i \left\vert \langle \psi(\tau) | k,i \rangle \right\vert^2,
  \qquad \text{with} \qquad |\psi(\tau) \rangle = e^{- i H_I \tau} | \psi(0) \rangle.
\end{equation}
Numerical results for the $p_k(\tau)$ probabilities displayed in Fig.~\ref{fig:SYK_QB}(a)
show that the occupation of the various states becomes uniform after a short transient time,
in which a collective charging rapidly sets in.
The levels are populated in a homogeneous and random way, as is typical for a chaotic and scrambling
model like the SYK one. In fact, it can be shown that the population $p_k$ of each single energy level is
proportional to its degeneracy, so that $p_k(\tau) \approx \frac{1}{2^N} \binom{N}{k}$.
In conclusion, the energy of the fermionic-SYK QB model preserves extensivity with $N$, such that $E^\#(\tau) \propto N$,
while the power behaves superextensively as
\begin{equation}
  P^\#_N(\tau) \propto N^{3/2} .
\end{equation}

We conclude by noticing that the enhanced performance of the SYK charging protocol over parallel charging,
or over more conventional QB models exhibiting extensivity in the power, is lost if
fermions in Eq.~\eqref{eq:SYK_ham} are replaced by bosons ($c_j \to b_j$).
In fact, the charging protocol for a modified bosonic version of the SYK interaction Hamiltonian
generates a dynamics that is clearly local in energy space, as reported in Fig.~\ref{fig:SYK_QB}(b).
More specifically, its poor performance
with respect to the fermionic SYK counterpart suggests that random pair hopping is not enough to warrant
superextensivity. On the other hand, non-local JW strings for fermions are crucial, as they maximize
entanglement production during the time evolution and thus correlations between the various quantum cells.

\section{Experimental platforms}
\label{experimental}

The quantum simulations presented in the previous sections can, in principle, be executed on any suitable experimental platform that can realize the required Hamiltonian evolution. In this section, we will briefly review what it takes to build such a quantum simulator or quantum computer. 

When Richard Feynman, at the Physics of Computation conference in 1981 mentioned in Sec.~\ref{sec:background}, first asked the question whether is was possible to simulate (quantum) nature on a classical computer, his answer was a clear ``no''~\cite{Feynman1982467}. Feynman realized that because of the exponential scaling of the size the Hilbert space with the number of constituents due to entanglement, simulating or calculating the quantum evolution of large systems ultimately make it necessary to use computers that also exploit the laws of quantum physics. Since then, building such computers has become not just an academic challenge, but also a commercial quest for ``supercomputers'' that could revolutionize many aspects of science and technology. 

The ultimate test of a quantum computer or simulator is the so-called quantum advantage (or quantum supremacy)~\cite{harrow17}: a proof that a quantum computer can solve a problem for which there is no efficient classical algorithm or, more simply, that a quantum computer can calculate something (that may or may not be practically useful) that a classical computer cannot calculate in a reasonable time and with reasonable resources (also called ``practical quantum advantage''~\cite{daley22}). At the time of writing of this review, no convincing claim of academically or commercially relevant quantum advantage has been made, and the jury is still out on the question of the ideal physical platform for quantum computation and simulation. A number of comprehensive overviews have been published in the past few years in which all the physical realizations tried to date are explained in great detail, and we refer the interested reader to those reviews for a detailed picture of the state-of-the art and future perspectives, both from a practical and fundamental point of view~\cite{kajer20,monroe21,fause24,rama23}. 

Generally speaking, to build a quantum computer (in the following by ``quantum computer'' we will always mean ``quantum computer or simulator'') one first needs to have a physical system in which one can clearly identify two (out of possibly many) quantum states that can be assigned the quantum-logical values $|0\rangle$  and $|1\rangle$  - the quantum bit or qubit. This is the first of the well-known DiVincenzo criteria~\cite{divin00}, which also includes the need for scalability - it must be feasible to have not just one, but many qubits in a quantum computer in order to perform complex tasks. Furthermore, it must be possible to initialize the qubits in some well-defined initial state, and also to implement a set of universal one- and two-qubit quantum gates in a time that is much shorter than the coherence time of the qubits. Finally, there needs to be a protocol for the read-out of the final state of the system. 

By now, several physical systems have been shown to satisfy the DiVincenzo criteria. These systems fall into two broad categories: natural and man-made. The natural qubit systems are atoms (or ions), molecules and photons, whereas the man-made ones include superconducting (Josephson junction) qubits, quantum dots and exotic materials that are believed to host Majorana fermions which feature topological protection against decoherence. 

Historically, the first demonstrations of quantum gates and quantum algorithms were achieved using trapped ions~\cite{monroe95} and molecules manipulated by NMR techniques~\cite{jones98}. While NMR-based quantum computation has essentially been abandoned because of the obvious problems regarding scalability, trapped ions continue to be intensively investigated as possible candidates for large-scale quantum computers, along with neutral Rydberg atoms, superconducting qubits and photons. In the following, we will give a concise overview of these four approaches.

\subsection{Trapped ions and Rydberg atoms}

The idea behind using atoms as physical systems for realizing qubits is simple: one identifies some suitable energy levels of the atoms that are used to represent $|0\rangle$ and $|1\rangle$, finds ways to induce a coupling between them in order to control the state of the qubit and to perform the read-out. Also, there has to be some interaction between the atoms that allows for the realization of two-qubit gates, and one has to be able hold the atoms in place for a sufficiently long time during which the gate operations or quantum evolution take place. 

One practical solution that satisfies these requirements is the use of ions trapped in electric traps~\cite{cirac95}, either of the Paul~\cite{holz02} or, more recently, the Penning type~\cite{jain24}. Once the ions are trapped, they are cooled down to micro-Kelvin temperatures using laser-cooling techniques. The electric charge on the ions naturally leads to a Coulomb interaction between them. This makes it possible to implement, for instance, a controlled-NOT quantum gate. The first demonstration of this gate was done by the group of David Wineland in 1995~\cite{monroe95}, where a single beryllium ion was used: the two qubits were represented by two internal states of the ion and its two lowest vibrational states. Gate operations based on the shared motional modes of two or more ions mediated by the Coulomb interaction, leading to entanglement between the ions, were demonstrated a few years later~\cite{Bruze19}. 

Depending on the energy levels chosen to define the qubit - hyperfine state, Zeeman sublevels, or states separated by energies corresponding to optical wavelengths - the coherence times of ion qubits range from hundreds of seconds to a fraction of a second. Even at the lower end, though, the coherence times are much longer than the times required to execute a one- or two-qubit gate, which are between a few microseconds and around a hundred microseconds~\cite{Bruze19}.

\begin{figure}
  \begin{centering}
    \includegraphics[width=1.\columnwidth]{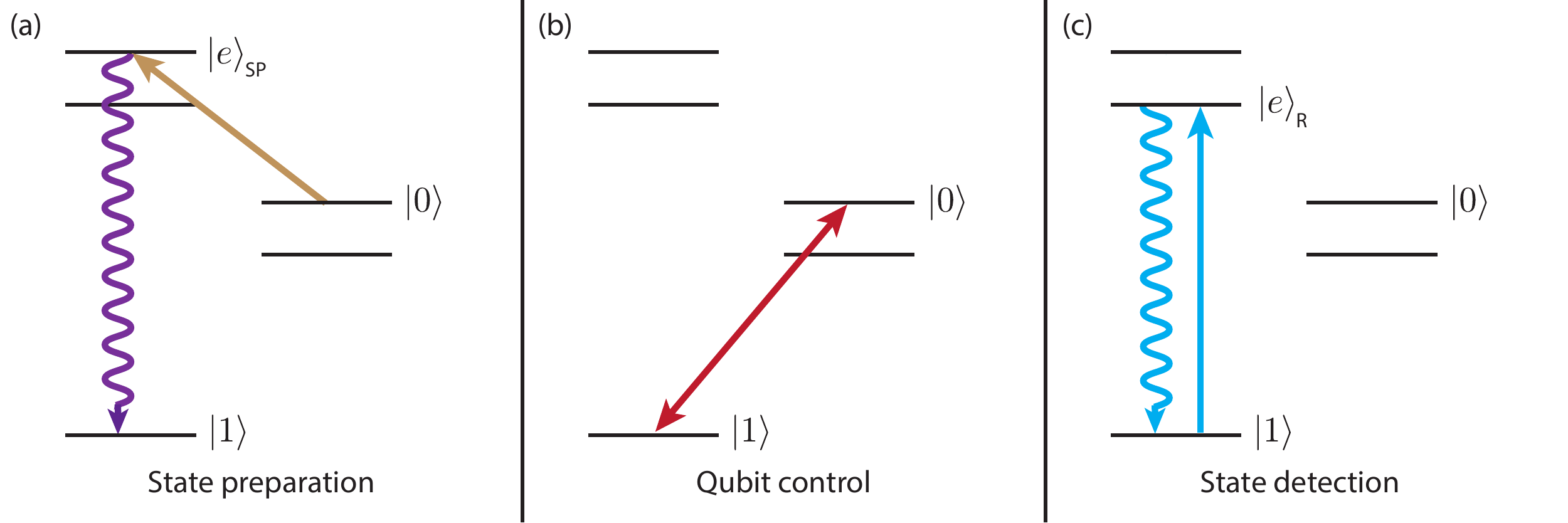}
    \caption{Ion trapped ion quantum computing, qubits can be initialized by optical pumping from a long-live state via a fast-decaying state (a). The state of the qubits can be manipulated via laser coupling (b), and read-out can be achieved by selectively coupling one of the qubit states to an excited state on observing the fluorescence from that state. From Ref.~\cite{Bruze19}.}
    \label{Fig_ioncontrol}
  \end{centering}
\end{figure}

Initially, the $|0\rangle$ or $|1\rangle$ states of the qubit is prepared via optical pumping, e.g., from some long-lived state to a short-lived state from which the ion can rapidly decay into the desired initial state (see Fig.~\ref{Fig_ioncontrol}). Readout of a trapped ion qubit is achieved by driving a transition that is resonant, e.g., with the $|1\rangle$ state of the qubit but not with the  $|0\rangle$ state and observing the fluorescence induced by the driving. When fluorescence is detected, one can infer that the qubit is in the  $|1\rangle$ state.

Some recent advances in ion trap quantum computing include the realization of a scalable version of Shor’s factoring algorithm~\cite{monz16} using 11 qubits and the demonstration of a racetrack ion trap architecture in which entangled states of 32 qubits could be realized~\cite{moses23}. Quantum simulations have also been carried out on trapped ion devices, such as the digital simulation of the Schwinger model (1+1 dimensional quantum electrodynamics)~\cite{martinez16}, and quantum simulations of high energy physics are also envisioned on such devices~\cite{bauer23}.

A more recent approach involving atomic qubits is based on Rydberg atoms~\cite{saff10,mor16} trapped in arrays of optical dipole traps, also called optical tweezers~\cite{lahaye22}. In a Rydberg atom, an outer electron is excited to a high-lying energy state with a principal quantum number $n> 10$. Such Rydberg states have much longer lifetimes than low-lying energy states. In alkali atoms, for example, these liftetimes can be several hundreds of microseconds as opposed to tens of nanoseconds. Rydberg atoms are (in general) charge-neutral, so in order to realize two-qubit gates, on needs an interaction that replaces the Coulomb repulsion of ions. That interaction is the van der Waals interaction between Rydberg atoms, which is much stronger than that between ground-state atoms and scales as $n^{11}$. Again, taking alkali atoms as an example, the van der Waals interaction between Rydberg atoms with a principal quantum number around $70$ at a distance of $10\,\mathrm{\mu m}$ can be tens of MHz, compared to a few milli-Hertz for the ground state. 

Neutral atoms can be trapped in free space using the ac-Stark shift (or light shift) of their energy levels induced by, e.g., a laser beam that is detuned with respect to an atomic transition~\cite{pesce20}. If the intensity of the laser varies in space, the resulting energy gradient gives rise to a force which, depending on the detuning, can either attract the atom to regions of higher intensity (negative or red detuning) or repel them from those regions (positive or blue detuning). For a focused, red-detuned Gaussian beam atoms are trapped around the focus of the beam.

Arrays of red-detuned optical dipole traps can be created using spatial light modulators, micromirror devices or acousto-optical deflectors (or a combination thereof)~\cite{morgado21}. The distance between the traps is typically a few micrometers. The traps are loaded from a cloud of cold atoms, typically a MOT. This loading process is random, and there will be a distribution of atom numbers in the individual traps depending on the density of the atomic cloud. If the waists of the dipole trap beams are on the order of $1\,\mathrm{\mu m}$, light-assisted collisions between atoms can lead to the occupation of a dipole trap by exactly one atom. Initially empty traps can be eliminated by re-arranging the traps in the array (the empty traps are identified by fluorescence imaging). 

Once a defect-free array of trapped atoms has been created, quantum operations can be carried out by using a combination of (possibly detuned) excitation lasers that couple the ground state of the atoms to a Rydberg state, additional focused laser beams that induce local ac Stark shifts, and microwave radiation. During this process, the dipole traps are temporarily switched off.

At the end of a computation, read-out is achieved by switching the dipole traps back on, thus re-capturing those atoms that are in the ground state. Atoms in the Rydberg state, by contrast, will not be re-captured as the dipole trap lasers do not induce an appropriate ac Stark shift on them (except in the special case of so-called magic wavelengths, which are typically not used in the experiments referred to here). After a few milliseconds, those atoms that are not trapped will fall down due to gravity and leave an empty trap behind. A resonant laser is now briefly flashed on, and the resulting fluorescence can be captured on a CCD to identify which traps are full and which ones are empty. 

\begin{figure}
  \begin{centering}
    \includegraphics[width=12 cm]{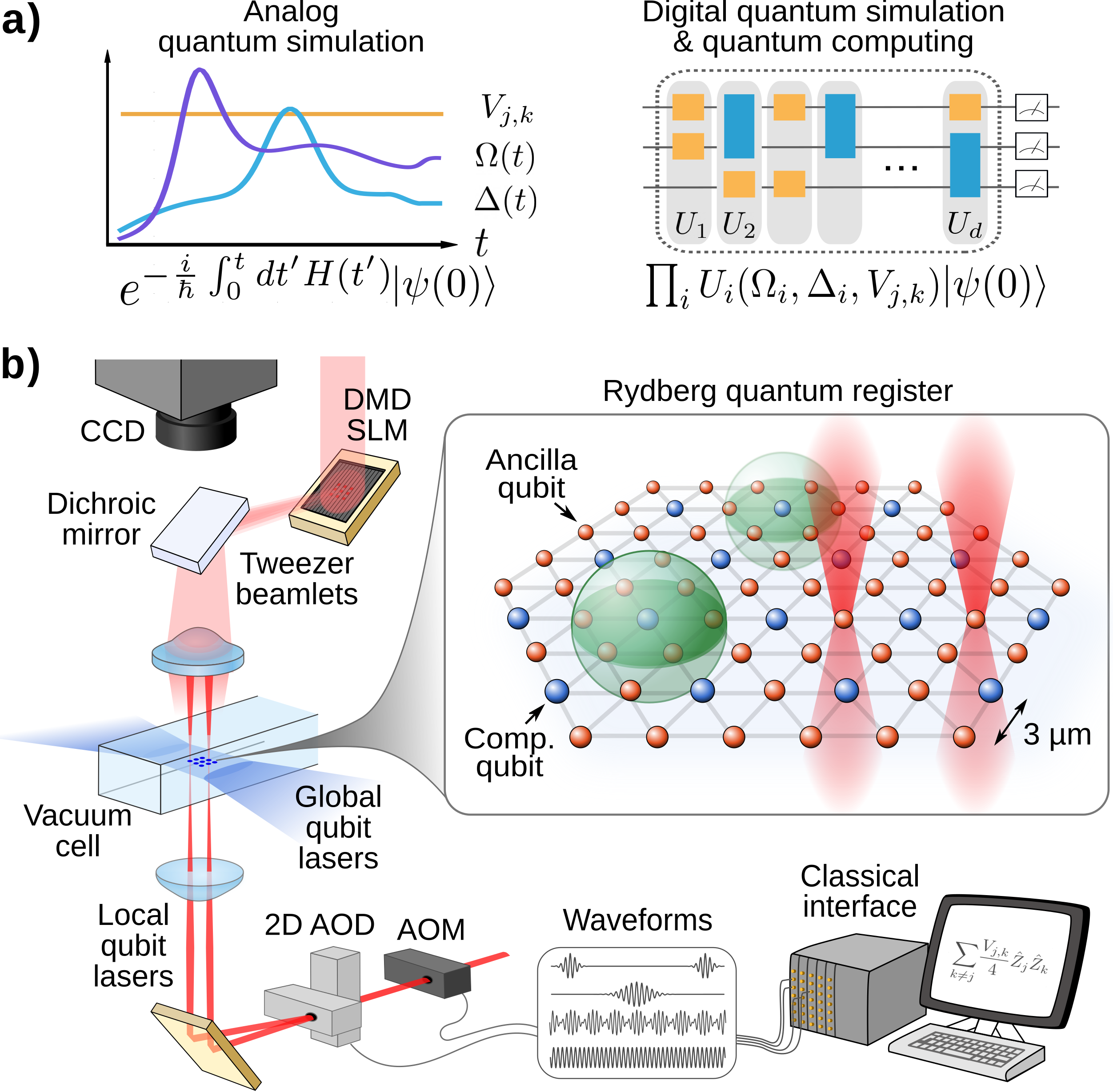}
    \caption{(a) An array of trapped Rydberg atoms can be used to carry out anologue or digital quantum simulations. (b) In a typical experimental setup, atoms are held in an array of optical dipole traps. Read-out is achieved by imaging the fluorescence of re-trapped atoms on a CCD camera. From Ref.~\cite{morgado21}.}
    \label{Fig_Rydberg}
  \end{centering}
\end{figure}

Arrays of trapped Rydberg atoms lend themselves particularly well to the realization of analogue quantum simulators. In such simulators, the evolution of a quantum system is studied by implementing a desired (possibly time-dependent) Hamiltonian $H(t)$ on the Rydberg array, choosing the geometry of the dipole traps and appropriate parameters for the driving. These parameters can also be varied in time, for instance by varying the driving strength or the detuning of the driving lasers from resonance. The Rydberg array effectively calculates the time evolution operator $\exp[{-\frac{i}{\hbar} \int_{0}^{t} dx H(t'))]}$.  After some evolution time, information on the system can be read out as described above. An example of a quantum simulation carried out on such a system is the recent work by the Lukin group~\cite{seme21}, in which a Rydberg quantum simulator with more than 200 qubits was used to probe quantum spin liquid states in a kagome lattice. 

Figure~\ref{Fig_Rydberg} shows a schematic view of a typical experimental setup for Rydberg quantum simulation. In such simulators it is also possible to implement digital quantum simulation, i.e., using the Lie-Trotter-Suzuki decomposition to break down the time evolution of the system into small steps, each represented by a unitary operator. This approach is equivalent to the execution of a quantum algorithm on a quantum computer.

\subsection{Superconducting platforms}

\begin{figure}
  \begin{centering}
    \includegraphics[width=12 cm]{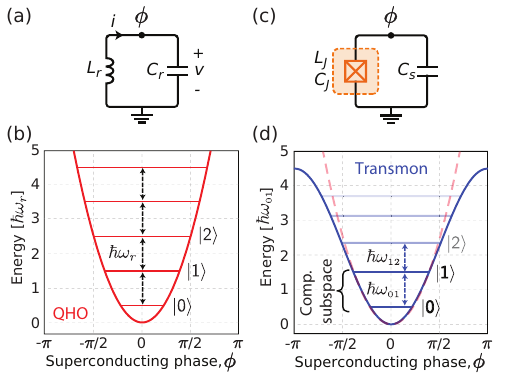}   
    \caption{The energy spectrum of a superconducting LC circuit consists of equally spaced levels (a). To be able to use such a circuit as a qubit, it is necessary to introduce a nonlinear element - the Josephson junction - into the circuit (b). This nonlinearity allows one to selectively drive transitions between the two lowest energy levels. From Ref.~\cite{krantz2019}.}
    \label{Fig_superconducting}
  \end{centering}
\end{figure}

When an electric LC circuit is cooled down sufficiently so that it becomes superconducting, the current oscillations in it can be quantized. The corresponding energy levels are those of a quantum harmonic oscillator and, therefore, evenly spaced (see Fig.~\ref{Fig_superconducting}). Such an energy spectrum does not lend itself immediately to the realization of a qubit, since any external driving needed to manipulate the qubit states would act on all pairs of adjacent energy levels in the same way. If the qubit is initially prepared in the lowest oscillatory state, an external drive can excite it not just to the first excited level, but successively to any higher energy level. 

To make superconducting circuits useful for quantum computation, it is necessary to introduce a nonlinear element into the circuit in order to make the energy spacing different between adjacent levels~\cite{krantz2019}. This nonlinear element is the Josephson junction, which consists of two superconductors separated by a thin insulating layer. Using a combination of inductances, capacitances, and Josephson junctions, it is possible to make superconducting qubits tailored specifically to certain applications, such as the transmon qubit which is now widely used~\cite{kajer20}. Single-qubit gates can be executed by coupling microwaves with appropriate frequencies and phases to the qubits, thus inducing Rabi oscillations (i.e., rotations on the Bloch sphere) around a desired axis. Two-qubit gates can be realized using a number of different techniques, such as parametric driving of a coupling element between two adjacent superconducting qubits that induces a tunable coupling between the qubits. 

Readout of superconducting qubits is achieved by coupling the qubit to a resonator whose resonant frequency is detuned with respect to that of the qubit. In this way, there is no energy exchange between the qubit and the resonator, but the state of the qubit can be inferred from the shift in the resonator resonance induced by it. This is also referred to as dispersive readout. 

Superconducting qubits can be manufactured using techniques that have been developed  for decades by the semiconductor industry, such as lithographic patterning, deposition, etching, and controlled oxidation~\cite{kajer20}. Several companies have adopted superconducting qubits as their platform of choice, such as IBM, Google, Rigetti and D-Wave. Since 2016, IBM offers a cloud service that allows users to run quantum algorithms on processors with a handful of qubits~\cite{ibm}. 

In general, quantum computing platforms based on superconducting qubits are designed to run quantum algorithms in terms of one- and two-qubit gates executed sequentially. This includes digital quantum simulation (see above), in which the evolution of a quantum system is realized by breaking the time evolution operator down into small parts. However, superconducting qubits can also be used for analogue quantum simulation. For instance, one-dimensional Bose-Hubbard chains have been experimentally studied~\cite{xu20} using a 12-qubit quantum processor. The Bose-Hubbard model was realized through quantum walks of photons on a one-dimensional array of 12 coupled superconducting transmon qubits~\cite{yan19}.

Another analogue approach to quantum computation is QA, in which the qubits and the couplings between them are tuned in such a way as to continuously vary the parameters of the Hamiltonian implemented by the qubits. The quantum computation then consists in performing that variation without exciting the system, and thus eventually finding its ground state. This approach has been followed mainly by D-Wave, which used QA on a 2000-qubit device to measure the phase diagram of a spin glass~\cite{harris18} and observe the Kosterlitz-Thouless phase transition on a frustrated Ising model~\cite{king18}.

\subsection{Photons}

\begin{figure}
  \begin{centering}
    \includegraphics[width=12 cm]{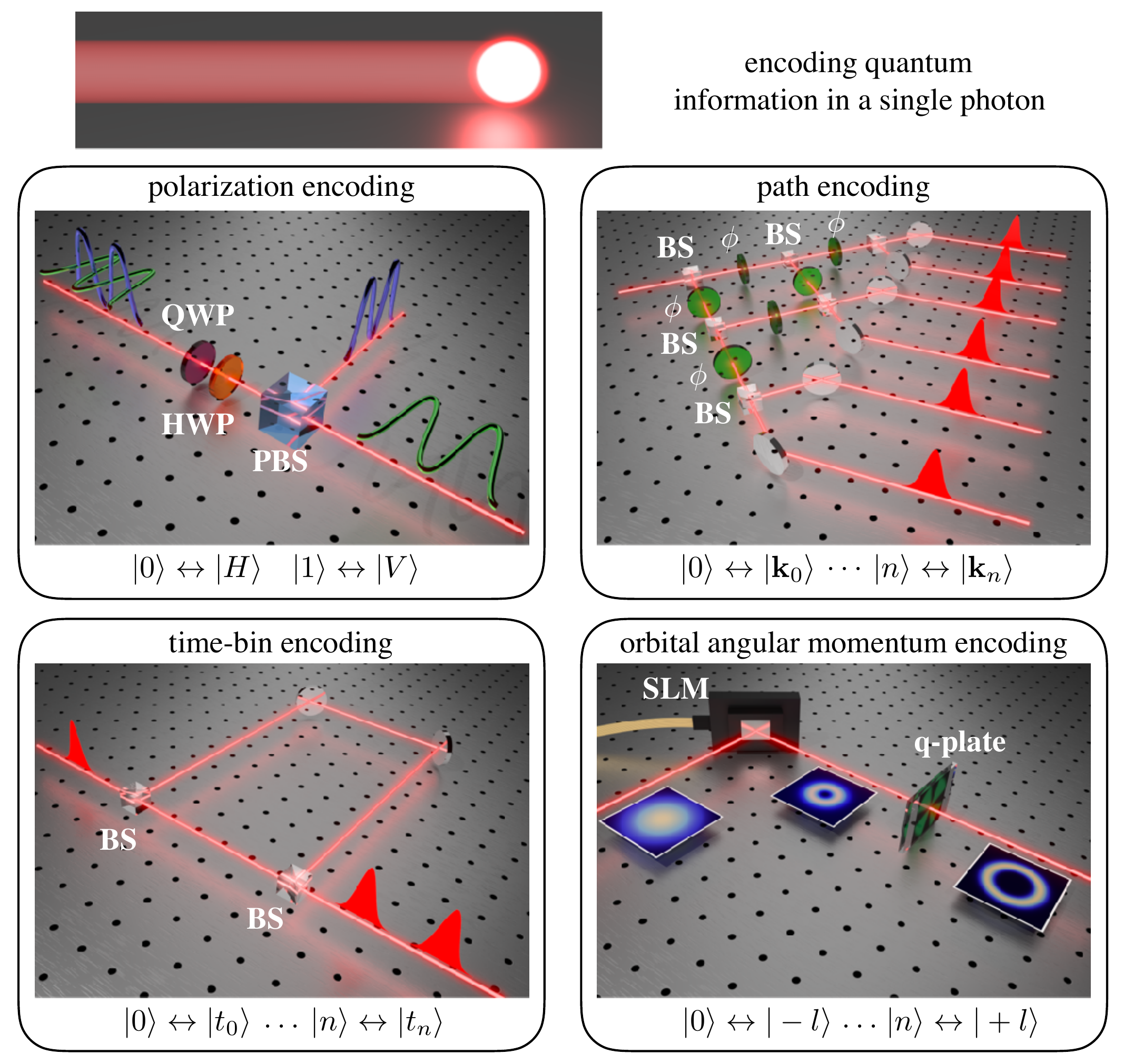}   
    \caption{Photons can be used as qubits using different types of encoding: polarization, path, time-bin, and orbital angular momentum. The state of the photon qubit can be controlled with waveplates (quarter-wave plate QWP, half-wave plate HWP), polarizing beam-splitters (PBS), beam splitters (BS), and q-plates (which couple the polarization of a photon to its orbital angular momentum). From Ref.~\cite{flam19}.}
    \label{Fig_photon_qubits}
  \end{centering}  
\end{figure}

Photons have been used extensively in quantum communication over the past forty years, for instance in quantum key exchange (quantum cryptography) and quantum teleportation. More recently, they have also been considered for applications in quantum simulation and computing~\cite{cout23}. Several schemes have been developed to encode quantum information in photons: through their path of propagation, polarization, orbital angular momentum, or time (via time-binning) (see Fig.~\ref{Fig_photon_qubits}). 

Path encoding amounts to assigning the qubit values $|0\rangle$ and $|1\rangle$ to specific directions of propagation, for instance through waveguides on a chip. Manipulation of the qubits is achieved by coupling the waveguides to each other, thus realizing beamsplitters. 

Polarization encoding associates the qubit values with, for instance, horizontal or vertical polarization of the photons $|H\rangle$ or $|V\rangle$. Equivalently, a rotated linear polarization basis can be chosen, such as $|\pm\rangle=\frac{1}{\sqrt{2}} (|H\rangle \pm |V\rangle)$, or a basis $|L/R\rangle=\frac{1}{\sqrt{2}} (|H\rangle \pm i |V\rangle)$
of left- or right-handed circularly polarized photons. The qubits in this representation can be manipulated using waveplates.

Orbital angular momentum (OAM) encoding is based on the spatial structure of the wavefront of the light, such as in Laguerre-Gauss beams, which have a phase profile corresponding to an optical vortex (or helical mode) with a phase term $e^{-iq\theta}$. Here, $\theta$ is the angular coordinate and $q$ is an integer describing the phase winding. A single photon has an OAM given by $L=q\hbar$. 

Finally, time-bin encoding is realized using a Mach-Zehnder interferometer with two arms of different lengths. The two qubit states then correspond to the photon having taken either the long or the short arm. 

An obvious difference between photonic qubits and the other physical realizations presented above is that photons do not interact with each other - unless, for instance, they propagate inside a nonlinear material. At first glance, this might seem to violate one of the DiVincenzo criteria, namely that a two-qubit gate is needed in order to have a universal set of quantum gates. However, in 2001 it was shown by Knill, Laflamme and Milburn~\cite{knill01} that universal quantum computing is possible using only linear operations involving beam splitters, phase shifters, single photon sources and photo-detectors. A few years later, Childs~\cite{childs09} demonstrated that any quantum computation can be recast in terms of a quantum walk~\cite{brom09}.

\subsection{Prospects for the future}

As mentioned in the introduction to this section, despite considerable progress has been made in quantum computing and quantum simulation over the past thirty years, there is no clear ``winner" among the physical platforms described above. All of them have their pros and cons: for instance, superconducting qubits achieve high gate fidelities but need sophisticated cryogenics. Trapped ions also achieve high gate fidelities but are not straightforward to scale to large numbers (hundreds or thousands) of qubits. Rydberg-atom based systems, on the other hand, are not currently suitable for gate-based quantum computing due to lower gate fidelity, but are being successfully used as quantum simulators with good scaling properties.

In terms of technological maturity and market-readiness, one might argue that quantum computers based on superconducting qubits and trapped ions as well as those based on Rydberg atoms are currently the frontrunners, with companies already selling products based on these technologies (as opposed to photonic or quantum dot approaches, which have not reached that level yet). However, the field is still very much evolving, as witnessed, for instance, by the recent announcement (February 2025) by Microsoft that they have realized a chip containing eight topological qubits~\cite{castelvecchi25,azurequantum25}. Such qubits, based on quasiparticles also known as Majorana fermions, are topologically protected from decoherence and could, therefore, be used to implement quantum algorithms without the need for error correction. In other platforms, error correction requires a considerable overhead of qubits with tens or hundreds of physical qubits per logical qubit. If the number of topological qubits could be scaled to the hundreds of thousands, this would give topological quantum computing a huge advantage over all other approaches. In the meantime, research on superconductors, ions etc. will no doubt continue and enable the realization on small-scale quantum simulators of some of the models discussed in this review.

\section{Summary}
\label{conclusions}

Guided by our feelings and interests, we have given a perspective on few modern aspects
of quantum simulation, involving the possibility to exploit peculiarities
of complex quantum many-body systems for quantum technologies.
We first contextualized our dissertation in the history of conventional quantum information processing
and quantum computing, whose origins can be dated back to the early 1980's, emphasising
some modern approaches such as quantum neural networks and reservoir computing.
As a concrete example of a simulatable complex system, we focused in some detail
on the Sachdev-Ye-Kitaev model, a paradigm to understand a variety of physical situations,
ranging from the behavior of strange metals (in condensed matter physics)
to that of black holes (in high energy physics).
We showed how the fast scrambling chaotic dynamics of the SYK model can be exploited to
boost the performances of the so-called quantum batteries, i.e., quantum devices that might be
capable to store energy more efficiently than with classical logic.
Finally, we provided a description of some promising experimental platforms to be realized in the lab.

\backmatter

\vspace*{5mm}

\bmhead{Acknowledgements}

G.M.P. thanks L. Innocenti, G. LoMonaco, S.Lorenzo, and M. Paternostro for a long lasting collaboration
on the topics of this review.
D.R. acknowledges the contribution of the collaborators involved on the topics here reviewed,
in particular G. M. Andolina, G. Pellitteri, and M. Polini.

\bmhead{Author contributions}
All the authors contributed to the conception, drafting, critical revision and literature search for this review article.

\bmhead{Funding}
G.M.P. acknowledges support by MUR under PRIN Project No. 2022 FEXLYB Quantum Reservoir Computing (QuReCo)
and by the ‘National Centre for HPC, Big Data and Quantum Computing (HPC)’ Project CN00000013 HyQELM-SPOKE 10.
O.M. acknowledges support by the Julian Schwinger Foundation through grant JSF-18-12-0011 and by MUR
(Ministero dell’Università e della Ricerca) through PNRR MUR project PE0000023-NQSTI.\\

\noindent
This version of the article has been accepted for publication, after peer review, but is not the Version of Record
and does not reflect post-acceptance improvements, or any corrections.
The Version of Record is available online at: \texttt{https://doi.org/10.1007/s40766-025-00069-0}

\bibliography{sn-bibliography}

\end{document}